\documentclass[%
reprint,
amsmath,amssymb,
aps,
pra,
]{revtex4-2}

\usepackage{graphicx}
\usepackage{bm}
\usepackage[hidelinks]{hyperref}
\DeclareMathOperator{\sinc}{sinc}
\DeclareMathOperator{\sgn}{sgn}
\DeclareMathOperator{\erf}{erf}
\DeclareMathOperator{\Si}{Si}

\begin{document}

\title{Interferometric sorting of temporal Hermite-Gauss modes via temporal Gouy phase}

\author{Dmitri B. Horoshko}\email{dmitri.horoshko@gmail.com}
\affiliation{Univ. Lille, CNRS, UMR 8523 - PhLAM - Physique des Lasers Atomes et Mol\'{e}cules, F-59000 Lille, France}
\author{Mikhail I. Kolobov}
\affiliation{Univ. Lille, CNRS, UMR 8523 - PhLAM - Physique des Lasers Atomes et Mol\'{e}cules, F-59000 Lille, France}

\date{\today}

\begin{abstract}
We propose a device consisting of $m$ Mach-Zehnder interferometers and realizing sorting of the first $2^m$ temporal Hermite-Gauss modes of light passing through it by adjusting the accumulated temporal Gouy phase acquired by every mode. This mode-order-dependent phase shift is achieved by a fractional Fourier transform realized by a time lens in one of the interferometer's arms. We consider application of such a sorter with just two interferometers to sorting the Schmidt modes of a photon pair generated in spontaneous parametric downconversion and find the theoretical lower bound on the cross-talk probability of 5.5\%. 
\end{abstract}

\maketitle

\section{Introduction} 
Decomposition of the electromagnetic field into a set of orthogonal spatio-temporal and polarization modes is a common practice in optics, plane monochromatic waves \cite{Dirac-Book} and resonator modes \cite{Siegman-Book} being the most notorious examples. Versatility of modal decompositions is determined by the fact that any complete orthonormal set of functions of spatial, temporal and polarization variables can be used as a modal basis for the field \cite{Fabre20}. In the last decade, much interest has been attracted to temporal modes of optical pulses, which provide a reliable platform for quantum information encoding ~\cite{Brecht15,Raymer20,Karpinski21} in numerous applications such as linear optics quantum computation \cite{Humphreys13,Lukens17}, boson sampling \cite{Shchesnovich14,Pant16}, quantum communications \cite{Nunn13,Islam17} and quantum sensing \cite{Jian12,Donohue18}. A key problem when using any modal basis for information encoding is an efficient mode sorting (demultiplexing), allowing one to address individually every mode and the inverse process of mode combining (multiplexing) for a compact joint transmission. 

Any polarization modal basis, intrinsically two-dimensional, can be rather easily sorted by a set of birefringent elements and polarizing beam splitters. Spatial modes are easily sorted only when they are non-overlapping in space (pixels) or in wavevector space (plane waves). For sorting overlapping spatial modes in free space, three groups of approaches can be distinguished.  First, overlapping modes can be mapped onto spatially separated pixels by interleaved application of spatial phase modulation and diffractive propagation, a technique known as multiple plane light conversion, which was successfully applied to Hermite-Gauss (HG) modes \cite{Morizur10,Labroille14,Fontaine19,Boucher20,Kupianskyi23}. Second, overlapping modes can be mapped onto quasi-plane waves. For Orbital Angular Momentum (OAM) modes it means a log-polar to Cartesian coordinate transform realized by spatial light modulators \cite{Berkhout10} or refractive optical elements \cite{Lavery12,Dudley13,Mirhosseini13}. Third, different spatial modes can be sent to different outputs of an interferometer when a mode-order-dependent phase shift is applied to the field in one of its arms. This approach was applied to sorting OAM modes \cite{Vasnetsov01,Leach02,Leach04,Ionicioiu16}, Laguerre–Gauss (LG) modes by the radial index \cite{Linares17a,Linares17b,Zhou17,Gu18}, and to free-space sorting of linear-polarized (LP) modes of few-mode fiber \cite{Igarashi15,Prieto-Blanco20}. The simplest variant of this approach, parity-sorting, is known in microscopy as image-inversion interferometry \cite{Sandeau06,Borges10,Nair16,Tang16}. Combinations of the second and third approaches were used for full sorting of LG \cite{Fu18} and HG \cite{Zhou18} modes by both indexes. In addition to these free-space sorters, a variety of spatial mode multiplexers exist for guided modes of optical fibers \cite{Li14}. 

In the temporal domain, modes are easily sorted if they are non-overlapping in time (time bins) or in frequency (frequency bins). Approaches to sorting overlapping temporal modes may be divided into the same three groups as for the spatial modes with the help of space-time analogy \cite{Akhmanov69,Kolner94,Patera18} based on the mathematical equivalence of descriptions of quadratic dispersion and paraxial diffraction. First, the overlapping modes can be mapped to temporally non-overlapping time bins by means of sequential temporal phase modulation and dispersion, which was shown for temporal HG modes~\cite{Ashby20,Joshi22}. Second, mapping temporal modes to non-overlapping spectral regions requires a nonlinear field transformation and can be achieved for temporal HG modes with a mode-selective frequency conversion by sum-frequency generation \cite{Eckstein11,Brecht14,Manurkar16,Shahverdi17,Ansari21} or four-wave mixing \cite{Reddy14,Reddy18}. Third, temporal modes can be parity sorted in a time-axis-inversion interferometer \cite{Mazelanik22}.

In this work, we present a method of sorting temporal Hermite-Gauss modes by means of arrays of interferometers with mode-order-dependent phase shift, providing thus an extension of the parity sorting of Ref.~\cite{Mazelanik22} to modulo-$2^m$ sorting, where $m$ is the number of used interferometers. Our approach shares the general idea with the approaches of the third group of spatial sorting, mentioned above.

In Sec. II, we review the general formalism for the description of the temporal degree of freedom in optics and introduce a temporal counterpart of the Gouy phase, which we call ``the temporal Gouy phase''. We also derive the temporal versions of the linear canonical transform and its particular case, the fractional Fourier transform, both well studied in the spatial domain \cite{Healy-Book}. In Sec. III, we describe the proposed temporal mode sorter for parity or modulo 4 sorting and show how it can be extended to modulo $2^m$ sorting. In Sec. IV, we consider an application of the proposed mode sorter to sorting the Schmidt modes of two entangled photons generated in spontaneous parametric downconversion (SPDC) and calculate analytically the cross-talk probability. The results are summarized in the Conclusion, Sec. V.

\section{General temporal formalism} 
\subsection{Temporal modes \label{sec:modes}}
Consider a discrete set of complex functions of time $\{\Psi_n(t),n=0,1,2...\}$ satisfying the conditions of orthonormality and completeness
\begin{eqnarray}\label{ortho}
\int \Psi_n(t)\Psi_m^*(t)dt &=& \delta_{nm}, \\\label{complete} 
\sum_{n=0}^\infty \Psi_n(t)\Psi_n^*(t') &=& \delta(t-t').
\end{eqnarray}
Here and below integration from $-\infty$ to $+\infty$ is implied if not otherwise specified. The set of Fourier-transformed functions 
\begin{equation}\label{Fourier}
\psi_n(\Omega) = \int \Psi_n(t)e^{i\Omega t}dt
\end{equation}
satisfies the conditions of orthonormality and completeness similar to Eqs.~(\ref{ortho}) and (\ref{complete}) with replacements $\Psi_n(t)\to\psi_n(\Omega)$, $dt\to d\Omega/2\pi$, $\delta(t-t')\to 2\pi\delta(\Omega-\Omega')$. 

The functions $\Psi_n(t)$ and $\psi_n(\Omega)$ can be considered as modal functions of the electromagnetic field in the time and frequency domains respectively. To show it, we consider propagation of an optical wave with the carrier frequency $\omega_0$  and carrier wavevector $k_0$ along the $z$ axis in a linear medium, where the wavevector can be approximated by a second-order polynomial of the frequency detuning $\Omega=\omega-\omega_0$, $k(\Omega)=k_0+k_0'\Omega+\frac12 k_0''\Omega^2$, $k_0'$ and $k_0''$ being the first and second derivatives of $k(\Omega)$ at $\Omega=0$. In such a medium, we write the positive-frequency part of electric field \cite{Mandel-Wolf} as
\begin{equation}\label{Efield}
E^{(+)}(z,t) = A(z,t-z/v_g)e^{ik_0z-i\omega_0t},
\end{equation}
where $A(z,t-z/v_g)$ is the envelope function at delayed time $t-z/v_g$,  $v_g=1/k'_0$ being the group velocity. When a wave passes through multiple media, the temporal argument should be delayed by the total group delay in all media \cite{Kolner94,Patera18}. Here, we do not consider the transverse coordinates of the optical wave implying they are decoupled from $z$ and $t$ during propagation through achromatic optical elements. The units for the electric field are chosen so that $E^{(-)}(z,t)E^{(+)}(z,t)$ represents the photon flux. Making a Fourier decomposition
\begin{equation}\label{Afield}
A(z,t) = \int a(z,\Omega)e^{-i\Omega t}\frac{d\Omega}{2\pi},
\end{equation}
we obtain $a(z,\Omega)$, the annihilation operator for a photon with frequency $\omega_0+\Omega$ at point $z$, satisfying the canonical equal-space commutation relations $[a(z,\Omega),a^\dagger(z,\Omega')]=2\pi\delta(\Omega-\Omega')$ \cite{Huttner90,Kolobov99}. In a non-dispersive medium $k''_0=0$ and the field propagation is described by a simple delay: $E^{(+)}(z,t)=E^{(+)}(z_0,t-[z-z_0]/v)$, where $v=\omega_0/k_0$ is the phase velocity, coinciding with the group one. It means that the time-delayed envelope remains invariant, $A(z,t)=A(z_0,t)$. Thus, it is sufficient to make a modal decomposition at some point $z_0$, to know it at any other point. Multiplying both sides of Eq.~(\ref{complete}) by $A(z_0,t')$ and integrating over $t'$, we obtain
\begin{equation}
A(z_0,t) = \sum_{n=0}^\infty A_n\Psi_n(t),
\end{equation}
where
\begin{equation}
A_n = \int A(z_0,t)\Psi_n^*(t)dt 
 = \int a(z_0,\Omega)\psi^*_n(\Omega)\frac{d\Omega}{2\pi}
\end{equation}
is the annihilation operator of a photon in the $n$th mode, satisfying $[A_n,A_m^\dagger]=\delta_{nm}$. 

\subsection{Dispersive medium and time lens}
In a dispersive medium, where $k''_0\ne0$, the frequency-domain operator acquires a phase shift when propagating over a distance $l$, $a(z+l,\Omega)=a(z,\Omega)\exp[ik''_0\Omega^2l/2]$, which means in the time domain  \cite{Patera18}
\begin{equation}\label{Dispersion}
A(z+l,t) = \int A(z,t')\frac{e^{-i(t-t')^2/2D}}{\sqrt{-2\pi iD}} dt',
\end{equation}
where $D=k_0''l$ is the group delay dispersion (GDD) of the medium. This transformation is a temporal counterpart of the Huygens-Fresnel integral for paraxial diffraction \cite{Siegman-Book}. The temporal modal functions undergo a similar integral transformation.

Another temporal counterpart of a spatial optical element is a time lens \cite{Kolner94,Bennett00a}, providing a quadratic temporal phase modulation of the signal pulse passing from $z'$ to $z$:
\begin{equation}\label{TimeLens}
A(z,t) = e^{it^2/2D_\mathrm{f}}A(z',t),
\end{equation}
where $D_\mathrm{f}$ is the focal GDD of the time lens. Time lenses are realized by electrooptical phase modulation or by nonlinear parametric processes such as sum-frequency generation or four-wave-mixing, see Refs.~\cite{Copmany11,Salem13} for reviews. Effective processing of quantum states requires that a time lens is lossless, which, for a parametric lens, means perfect phase matching and unit quantum conversion efficiency \cite{Patera17,Shi20,Srivastava23,Srivastava23b}. Single photons were processed by time lenses based on sum-frequency generation \cite{Lavoie13}, four-wave mixing \cite{Joshi22}, electrooptical phase modulation \cite{Karpinski17,Mittal17,Sosnicki20,Sosnicki23}, and atomic-cloud-based quantum memory \cite{Mazelanik22}.

\subsection{Linear canonical transform} 
Similar to how it is done in the spatial domain, the field transformations, described by Eqs. (\ref{Dispersion}) and (\ref{TimeLens}) can be viewed as particular cases of a class of linear canonical transforms \cite{Healy-Book} having the form 
\begin{equation}\label{transform}
A(z,t) = \int K_\mathbf{T}(t,t')A(z',t')dt',
\end{equation}
where the kernel is 
\begin{equation}\label{KT}
K_\mathbf{T}(t,t') = \frac1{\sqrt{2\pi ib}}
\exp\left(\frac{i}{2b}\left[dt^2-2tt'+a{t'}^2\right]\right),
\end{equation}
which represents a normalized phase shift by a quadratic form of the variables $t$ and $t'$. The parametrization of this quadratic form by three real numbers $a$, $b$ and $d$ has a remarkable property: When they are supplemented by the fourth real number $c$ such that $ad-bc=1$, and written in the matrix form
\begin{equation}\label{Tmatrix}
\mathbf{T} = \left(
\begin{array}{cc}
a & b \\
c & d
\end{array}
\right),
\end{equation}
the transform has the property of cascadability, i.e., two consecutive transforms with kernels $K_{\mathbf{T}_1}(t,t')$ and $K_{\mathbf{T}_2}(t,t')$ create a transform of the same class with a kernel 
\begin{equation}
K_{\mathbf{T}}(t,t') = \int K_{\mathbf{T}_2}(t,t'') K_{\mathbf{T}_1}(t'',t')dt'', 
\end{equation}
where $\mathbf{T}=\mathbf{T}_2\mathbf{T}_1$. In the spatial domain, matrix $\mathbf{T}$ is a scaled version of the ray matrix, widely used in geometrical optics \cite{Siegman-Book}. For this reason, we call it ``temporal ray matrix''. Note that temporal ABCD matrices were introduced previously \cite{Nakazawa98} for the description of propagation of an optical pulse with a Gaussian temporal shape through modulators and dispersive media. The phases of coefficients $b$ and $c$ are shifted in this approach with respect to its spatial counterpart. In contrast, our formalism introduces no phase shifts and is applicable to any temporal shape of the propagating pulse. 

Comparing Eq. (\ref{Dispersion}) to Eq. (\ref{transform}), we find that dispersive propagation corresponds to the choice 
\begin{equation}\label{Tprop}
\mathbf{T}_\mathrm{prop}(D) =  \left(
\begin{array}{cc}
 1 & -D \\
 0 & 1
\end{array}
\right).
\end{equation}
In the limit $b\to0$, we consider $a$ and $c$ as independent variables and write $d=1/a+bc/a$. Substituting this into Eq. (\ref{KT}) and taking the limit, we obtain
\begin{equation}\label{transformb0}
K_\mathbf{T}(t,t') = \sqrt{a} e^{ic t^2/2a} \delta(t-at'),
\end{equation}
where we have used the relation $(i\pi\epsilon)^{-1/2}\exp\left(it^2/\epsilon\right) \to\delta(t)$ at $\epsilon\to0$, which can be verified by taking Fourier transforms of both its sides.
Comparing Eq. (\ref{transformb0}) to Eq.~(\ref{TimeLens}) we conclude that 
a time lens corresponds to  
\begin{equation}
\mathbf{T}_\mathrm{lens}(D_\mathrm{f}) =  \left(
 \begin{array}{cc}
 1 & 0 \\
 1/D_\mathrm{f} & 1
\end{array}
\right).
\end{equation}
Due to cascadability, any sequence of dispersive media and time lenses can be described by the product of the corresponding temporal ray matrices.

\subsection{Temporal Hermite-Gauss modes \label{sec:HGmodes}}
Among infinitely many choices for modal functions, we choose $\Psi_n(t)=h_n(t/\tau)/\sqrt{\tau}$, where 
$h_n(x) = \left(2^nn!\sqrt{\pi}\right)^{-\frac12}H_n(x)e^{-x^2/2}$ is the Hermite-Gauss function, $H_n(x)$ being the Hermite polynomial. The temporal width of the modal function is determined by the parameter $\tau$, having the dimension of time. These modes are a good approximation for the Schmidt modes of photon pairs generated in spontaneous parametric downconversion (SPDC), one of the key sources of quantum light in modern quantum photonics \cite{Brecht15,Ansari18}.

Transformation of $\Psi_n(t)$ when propagating through an optical system with arbitrary temporal ray matrix $\mathbf{T}$ can be obtained by adopting the dimensionless formula of Ref.~\cite{Healy-Book} to the temporal domain or by direct integration of the generating function for the Hermite polynomials (see Appendix~\ref{sec:appendix0}):
\begin{equation}\label{HGabcd}
\frac1{\sqrt{\tau}}h_n\left(\frac{t}{\tau}\right) \to
\frac{e^{-i\gamma(n+1/2)}}{\sqrt{\tau\beta}}
h_n\left(\frac{t}{\tau\beta}\right) 
e^{-\alpha (t/\tau)^2/2},
\end{equation}
where $\alpha=\left[(d-ic\tau^2)(a-ib/\tau^2)-1\right]/\beta^2$, $\beta=|a+ib/\tau^2|$, $\gamma=\arg(a+ib/\tau^2)$. In particular, propagation through a dispersive medium with the temporal ray matrix (\ref{Tprop}) increases the temporal width to $\tau_D=\tau\beta=\tau\sqrt{1+D^2/\tau^4}$. 

\subsection{Accumulated temporal Gouy phase}
The phase $\gamma$ may be rewritten in the general case as
\begin{equation}\label{gamma}
\gamma = \arctan\left(\frac{b}{a\tau^2}\right) + \pi\sgn(b)\theta(-a)
\end{equation}
with $\theta(x)$ the step function and $\arctan(x)\in[-\pi/2,\pi/2]$. This phase is known in the spatial domain as accumulated Gouy phase shift \cite{Erden97}. To justify this name in the temporal domain, let us consider the temporal counterpart of the Gouy phase shift.

In the spatial domain, a monochromatic field with a wavevector $k$ directed along the $z$ axis and a Gaussian transverse distribution at plane $z=0$, $E^{(+)}(x,y,0)=E_0\exp\left[-(x^2+y^2)/w_0^2\right]$, propagates diffractively so that at any other plane $z>0$ its transverse distribution, in the paraxial approximation, is \cite{Siegman-Book}
\begin{eqnarray}\label{GaussianBeam}
E^{(+)}(x,y,z)&=&E_0\frac{w_0}{w(z)}
\exp\left(i[kz-\Phi(z)]\right)\\\nonumber
&\times& \exp\left(-\frac{x^2+y^2}{w^2(z)} +ik\frac{x^2+y^2}{2R(z)}\right),
\end{eqnarray}
where $w(z)=w_0\sqrt{1+4z^2/(k^2w_0^4)}$ is the spot size, $R(z)=z[1+k^2w_0^4/(4z^2)]$ is the radius of curvature, and $\Phi(z)=\arctan\left(2z/kw_0^2\right)$ is the axial phase shift. Equation (\ref{GaussianBeam}) describes a divergent Gaussian beam. This equation is also valid for $z<0$ and describes a convergent Gaussian beam: If a curved wavefront is created for a Gaussian beam at a plane $z<0$, e.g. by a lens, the beam converges to its focus at $z=0$. A beam passing through the focus acquires between the planes $-z_0$ and $z_0$, where $z_0\gg kw_0^2/2$, a total axial ($x=y=0$) phase shift with respect to the phase of a plane wave $\Phi_\text{Gouy}=\Phi(z_0)-\Phi(-z_0)=\pi$, which is known as Gouy phase shift \cite{Siegman-Book}. 
\begin{figure}[!ht]
\centering
\includegraphics[width=\linewidth]{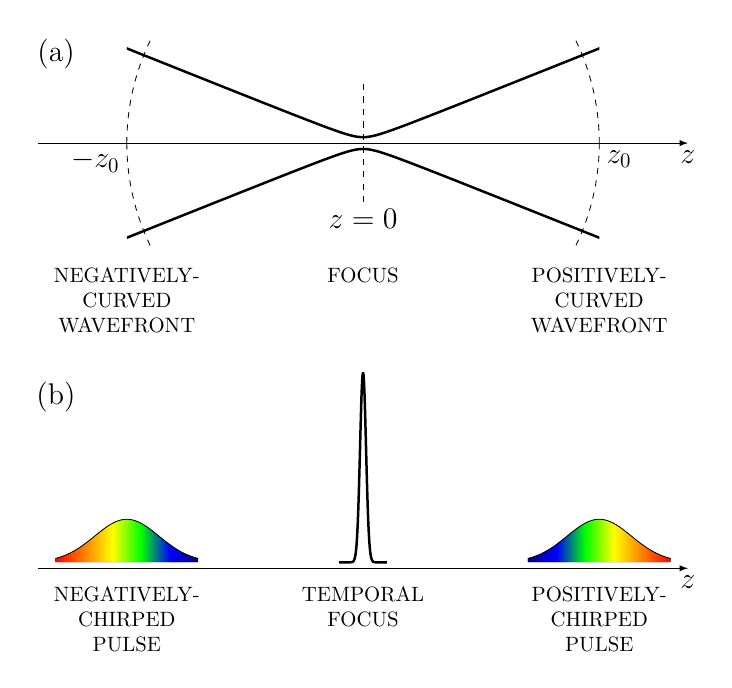}
\caption{(a) Focusing of a Gaussian beam. Thick solid lines show the beam radius $\pm w(z)$ at different planes, dashed lines show the wavefronts. (b) Temporal focusing of a Gaussian pulse in a medium with a positive dispersion ($k_0''>0$). In a negatively-chirped pulse, the blue part advances the red part. Since the group velocity for the blue part is lower than that for the red part, all spectral parts arrive at the temporal focus simultaneously. Then the red part advances the blue one resulting in a positively-chirped pulse.  \label{fig:Gouy}}
\end{figure}

In the temporal domain, a plane wave with central frequency $\omega_0$ and a Gaussian temporal distribution at point $z=0$ of its envelope, $A(0,t)=E_0\exp\left[-t^2/2\tau_0^2\right]$, propagates in a dispersive medium with the dispersion law $k(\Omega)$ so that at any other plane $z>0$ its temporal distribution, in the quadratic-dispersion approximation, is
\begin{eqnarray}\label{GaussianPulse}
    A(z,t)&=&E_0\sqrt{\frac{\tau_0}{\tau(z)}}
    \exp\left[-i\Phi^\text{temp}(z)\right]\\\nonumber
    &\times& \exp\left(-\frac{t^2}{2\tau^2(z)}-i\frac{t^2}{2R^\text{temp}(z)}\right)
\end{eqnarray}
where $\tau(z)=\tau_0\sqrt{1+D^2/\tau_0^4}$ is the temporal standard deviation, $R^\text{temp}(z)=D[1+\tau_0^4/D^2]$ is the temporal chirp rate, $\Phi^\text{temp}(z)=-\frac12\arctan\left(D/\tau_0^2\right)$ is the temporal-axial phase shift, and $D=k_0''z$ is the GDD between the two planes with $k_0''>0$. Equation (\ref{GaussianPulse}) is obtained by substituting the Gaussian envelope $A(0,t)$ specified above into Eq. (\ref{Dispersion}) and describes a temporally spreading Gaussian pulse. This equation is also valid for $z<0$ and describes a temporally focused Gaussian pulse: If a negative chirp rate is created for a Gaussian pulse at a plane $z<0$, e.g. by a time lens, the pulse temporally narrows to its transform-limited shape at $z=0$. A pulse passing through the temporal focus acquires between the planes $-z_0$ and $z_0$, where $k_0''z_0\gg \tau_0^2$, a total temporal-axial ($t=0$) phase shift $\Phi_\text{Gouy}^\text{temp}=\Phi^\text{temp}(z_0)-\Phi^\text{temp}(-z_0)=-\pi/2$, which can be called ``temporal Gouy phase shift.'' 

It should be noted that the temporal axis at plane $z$ means $t=0$ for the time-delayed envelope, and $t=z/v_g$ for the full field amplitude in accord with Eq. (\ref{Efield}). It means that the full phase of the field at the temporal axis at plane $z$ is $k_0z-\omega_0z/v_g-\Phi^\text{temp}(z)$ and the temporal-axial phase shift is defined with respect to the phase of a plane monochromatic wave delayed by the group velocity delay time $z/v_g$. This is a slight but important difference with the definition of the Gouy phase shift in the spatial domain.

We see that the temporal Gouy phase shift differs from its spatial counterpart in two aspects: It has the opposite sign and its modulus is twice less. The sign inversion is explained by different signs for spatial and temporal coordinates in the argument of a retarded-potential solution of the wave equation. The twice higher modulus in the spatial domain is explained by the focusing happening in two dimensions $x$ and $y$, while the temporal focusing is one-dimensional.

Propagation of a general Hermite-Gauss mode in a dispersive medium is described by Eq. (\ref{HGabcd}) with $\tau=\tau_0$ and the parameters of the temporal ray matrix taken from Eq. (\ref{Tprop}), i.e., $\alpha=|1-iD/\tau_0^2|=\tau(z)/\tau_0$, $\beta=iD/\tau^2(z)$, and $\gamma=-\arctan(D/\tau_0^2)$. For a Gaussian mode, $n=0$, we identify $\gamma/2$ as the temporal-axis phase shift discussed above, leading to the temporal Gouy phase shift in the temporal far-field limit. For higher-order modes the corresponding phase shift is $(n+\frac12)\gamma$. Note that for propagation in a system with a negative parameter $a$, $\gamma$ can take values in  $[-\pi,\pi]$ outside the range $[-\pi/2,\pi/2]$ given by the $\arctan(x)$ function. Similar to the spatial domain terminology, the phase shift  $(n+\frac12)\gamma$ will be called ``accumulated temporal Gouy phase.''

Note that recently a temporal counterpart of Gouy phase has been suggested for the construction of ``temporal cavities'' with a different expression for the mode-dependent phase shift \cite{Dioum23}. As indicated above, the Gouy phase shift requires a modification when passing from the spatial to the temporal domain.

\subsection{Fractional Fourier transform}
We see from Eq.~(\ref{HGabcd}) that Hermite-Gauss functions are eigenfunctions of a kernel with $\alpha=0$ and $\beta=1$, which is equivalent to two conditions $d=a$, $c\tau^2=-b/\tau^2$, wherefrom it follows that $|a+ib/\tau^2|=1$ (we recall that for any temporal ray matrix $ad-bc=1$). Such a kernel, additionally multiplied by $\exp(i\gamma/2)$, corresponds to a transformation known as fractional Fourier transform \cite{Healy-Book} realized recently in the temporal domain \cite{Niewelt23,Lipka24}. It is characterized by just one parameter, for which we choose $\gamma$ known in this context as ``fractional angle''. Substituting the required values of the parameters into Eq.~(\ref{Tmatrix}) we obtain the temporal ray matrix 
\begin{equation}\label{Tfract}
    \mathbf{T} = \left(
    \begin{array}{cc}
        \cos\gamma & \tau^2\sin\gamma \\
        -\frac1{\tau^2}\sin\gamma & \cos\gamma
    \end{array}
    \right),
\end{equation}
which corresponds to a fractional Fourier transform of order $2\gamma/\pi$ for a pulse of width $\tau$. The order shows the fraction of the ordinary Fourier transform which is realized, order 1 corresponding to the ordinary Fourier transform. Note, that in quantum optics, one decomposes the positive-frequency part of the field as function of time $t$ and position $z$ into plane monochromatic waves of the form $\exp\left(ikz-i\omega t\right)$, where $k$ and $\omega$ are the wave vector and circular frequency respectively, which is considered as an inverse Fourier transform in space and time \cite{Mandel-Wolf}. Thus, mathematically, the Fourier transform in time is similar to the inverse Fourier transform in space, hence the minus sign in the definition of the fractional order with respect to Ref. \cite{Healy-Book}. 

Fractional Fourier transform defined by Eq. (\ref{Tfract}) does not change the temporal width of the Hermite-Gauss mode if it is exactly $\tau$. A more general fractional Fourier transform can include a magnification, as the ordinary Fourier transform does, but in the method described below, the transformed pulse will interfere with the initial pulse, and for a perfect constructive or destructive interference, their widths should be the same. The field transformation by the fractional Fourier transform is given by Eq. (\ref{transform}) with the kernel
\begin{equation}\label{KTFract}
    \tilde K_\mathbf{T}(t,t') = 
    \frac{e^{i\gamma/2 +i\left(t^2\cos\gamma -2tt'+{t'}^2\cos\gamma\right)/\left(2\tau^2\sin\gamma\right)}}
    {\sqrt{2\pi i\tau^2\sin\gamma}},
\end{equation}
where the tilde denotes a kernel phase-shifted by $\gamma/2$ with respect to one defined by the matrix $\mathbf{T}$. In the limit $\sin\gamma\to0$, which happens at $\gamma=\pi m$ with an integer $m$, this kernel, according to Eq. (\ref{transformb0}), tends to
\begin{equation}\label{transformb0Fract}
   \tilde K_\mathbf{T}(t,t') = e^{i\gamma/2}\sqrt{(-1)^m}\delta[t-(-1)^mt'],
\end{equation}
where we have substituted $\cos(\pi m)=(-1)^m$. 

Type-1 canonical decomposition of linear canonical transform \cite{Healy-Book} suggests that a fractional Fourier transform can be realized by just one time lens of focal GDD $D_\mathrm{f}$ preceded and followed by dispersive elements of GDD $D$ each, supplemented by a phase shift $\gamma/2$. Such a system is described by the temporal ray matrix
\begin{eqnarray}\label{PLP}
   \mathbf{T}_\mathrm{prop}(D)\mathbf{T}_\mathrm{lens}(D_\mathrm{f}) \mathbf{T}_\mathrm{prop}(D)
   \phantom{TTTTTTTTTTTTT}\\\nonumber
   \phantom{TTTTTTTT} = \left(
    \begin{array}{cc}
        1-D/D_\mathrm{f} & -D(2-D/D_\mathrm{f}) \\
        1/D_\mathrm{f} & 1-D/D_\mathrm{f}
    \end{array}
    \right).
\end{eqnarray}
In order that this matrix coincides with that of Eq.~(\ref{Tfract}), two conditions should be satisfied
\begin{eqnarray}\label{Dcondition}
    &&D = D_\mathrm{f}(1-\cos\gamma),\\\label{DFcondition}
    &&D_\mathrm{f}\sin\gamma = -\tau^2.
\end{eqnarray}

Equation (\ref{HGabcd}) takes for a fractional Fourier transform a simple form 
\begin{equation}\label{HGabcdFrac}
    \frac1{\sqrt{\tau}}h_n\left(\frac{t}{\tau}\right) \to
    e^{-i\gamma n}\frac1{\sqrt{\tau}}
    h_n\left(\frac{t}{\tau}\right),
\end{equation}
i.e. a Hermite-Gauss function of order $n$ is an eigenfunction of the fractional Fourier transform with the eigenvalue $e^{-i\gamma n}$. When the input pulse contains exactly one photon, its quantum state can be considered as a superposition of an infinite number of basis states, corresponding to the photon occupying one of the Hermite-Gauss modes. In the Hilbert space spanned by such states, the fractional Fourier transform of fractional angle $\gamma=-2\pi/k$, $k=2,3,...$, realizes a gate, multiplying the $n$th basis state by $e^{i2\pi n/k}$, which is known as $Z_k$ gate \cite{Gottesman99}.

We see from Eq. (\ref{HGabcdFrac}), that the fractional Fourier transform does not change the bandwidth of a Hermite-Gauss mode with the temporal scale $\tau$. At first glance, this might seem unusual, because the quadratic phase modulation realized by the time lens changes the instantaneous frequency and, in general, the bandwidth of the field. However, in a fractional Fourier transform, satisfying Eqs. (\ref{Dcondition}) and (\ref{DFcondition}) the interplay of dispersion and phase modulation results in preservation of the bandwidth for any fractional angle. This can be seen explicitly for the zeroth-order mode (a Gaussian pulse) by calculating its bandwidth after a dispersive medium and a time lens and verifying that it is preserved when conditions (\ref{Dcondition}) and (\ref{DFcondition}) are satisfied (see Appendix \ref{sec:appendix1}).

\section{Temporal mode sorter} 
\begin{figure*}[ht]
\centering
\includegraphics[width=\linewidth]{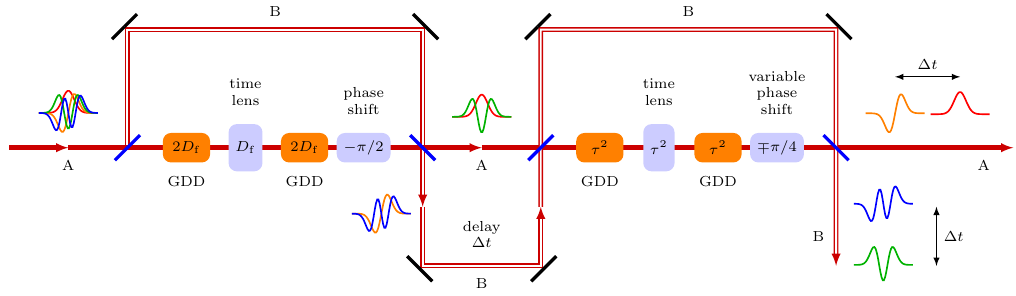}
\caption{Schematic of interferometric temporal mode sorter including just two interferometers $\ell=1$ and $\ell=2$ for sorting four first temporal Hermite-Gauss modes. Two arms of each interferometer are denoted by A (single line) and B (double line), which corresponds also to the notation of beams outside the interferometers. The arm A of each interferometer contains two dispersive elements (GDD), a time lens, and a phase shifter. At the sorter output, the modes of orders 0 and 1 appear in the beam A separated by time $\Delta t$ surpassing their durations. The modes of orders 2 and 3 appear in the beam B separated by the same time interval.  \label{fig:interf}}
\end{figure*}

The proposed temporal mode sorter is shown in Fig.~\ref{fig:interf}. It represents a cascade of Mach-Zehnder interferometers numbered by $\ell=1,2,...$, each containing in one of its arms a $Z_{2^\ell}$ gate realized by a single-time-lens fractional Fourier transform discussed above. 
The first beam splitter in every interferometer is symmetric and provides a linear transformation of its input fields written as a column vector $\mathbf{A}_\ell^\text{in} = [A_\ell^\text{in}(t),B_\ell^\text{in}(t)]^T$ to its output fields (inside the interferometer) $\mathbf{A}_\ell' = [A_\ell'(t),B_\ell'(t)]^T$  so that $\mathbf{A}'_\ell=\mathbb{U}\mathbf{A}_\ell^\text{in}$, where 
\begin{equation}\label{U}
    \mathbb{U} = \frac1{\sqrt{2}}\left(
    \begin{array}{cc}
        1 & 1 \\
        -1 & 1
    \end{array}
    \right)
\end{equation}
is the unitary transformation matrix of a symmetric beam splitter.

Both optical paths between the beam splitters in an interferometer should have (i) the same total optical phase shift at the carrier frequency $\omega_0$ and (ii) the same group delay of the field envelope, in order to produce constructive and destructive interference at the second beam splitter's output. These two conditions cannot be satisfied simultaneously by varying just one parameter, the length difference of the interferometer arms. Both conditions can be satisfied, e.g., by placing a dispersive element with the dispersion law $k_b(\Omega)$ into the arm B of the interferometer (containing no gate). If the total phase shift at carrier frequency and total group delay in the arm A (containing a $Z_{2^\ell}$ gate) are $\phi_a$ and $\tau_a$ respectively, then the length of the air part of the arm B $l_1$ and the length of the dispersive element $l_2$ should satisfy two equations: $\omega_0l_1/c+k_bl_2=\phi_a$ and $l_1/c+k_b'l_2=\tau_a$, where $k_b$ and $k_b'$ are coefficients of decomposition $k_b(\Omega)\approx k_b+k_b'\Omega$.

The second beam splitter in every interferometer will be described by matrix $\mathbb{U}^\dagger$ so that the total field transformation in an empty interferometer is identity: $\mathbb{U}^\dagger\mathbb{U}=\mathbb{I}$, where $\mathbb{I}$ is the unit $2\times2$ matrix. We denote the action of a $Z_k$ gate by operator $\mathcal{Z}_k$ so that the input field $A(t)$ is transformed into $\mathcal{Z}_kA(t)$ and write the field at the interferometer output as
\begin{eqnarray}\label{Aout}
    \mathbf{A}_\ell^\text{out} &=& \mathbb{U}^\dagger
    \left(
    \begin{array}{cc}
        e^{i\theta_\ell}\mathcal{Z}_{2^\ell} & 0 \\
        0 & 1
    \end{array}
    \right)\mathbb{U}\mathbf{A}_\ell^\text{in}\\\nonumber
    &=& \frac12\left(
    \begin{array}{cc}
        e^{i\theta_\ell}\mathcal{Z}_{2^\ell}+1 
        & e^{i\theta_\ell}\mathcal{Z}_{2^\ell}-1 \\
        e^{i\theta_\ell}\mathcal{Z}_{2^\ell}-1 
        & e^{i\theta_\ell}\mathcal{Z}_{2^\ell}+1
    \end{array}
    \right)\mathbf{A}_\ell^\text{in},
\end{eqnarray}
where $\theta_\ell$ is an additional phase shift necessary for reaching constructive and destructive interference at the interferometer output for the targeted temporal modes.

\subsection{Image inversion interferometer} 
The purpose of the first interferometer is to separate the even and odd order modes at two of its outputs. For this purpose, the interferometer is equipped with a $Z_2$ gate (image inverter) in its arm A, a device realising a temporal inversion of the input waveform. For this device the fractional angle $\gamma\to-\pi$ and from Eqs.~(\ref{Dcondition}) and (\ref{DFcondition}) we have $D = 2D_\mathrm{f}$ and $\tau^2/D_\mathrm{f}\ll1$, which corresponds to a temporal counterpart of the ``2f-2f'' spatial imaging system, having the magnification $-1$. Substituting these values to Eq. (\ref{PLP}), we obtain $a=d=-1$, $b=0$, $c=1/D_\mathrm{f}$ and from Eq. (\ref{transformb0}) we have $A(z,t) = iA(z',-t)\exp(-it^2/2D_\mathrm{f})$, where the time inversion is accompanied by a chirp and a $\pi/2$ phase shift (see also Ref. \cite{Patera18} for field transformation in a system with magnification $-1$). In the temporal far-field limit, required by the second condition, the chirp is negligible. The $\pi/2$ phase shift is compensated by the $\gamma/2=-\pi/2$ phase shift, which one needs to add to a linear canonical transform in order to obtain a fractional Fourier transform. The overall field transformation is
\begin{equation}\label{transformInv}
   A(z,t) = \mathcal{Z}_2A(z',t) = A(z',-t). 
\end{equation}
If a long dispersive medium is not desirable, the chirp can be removed by a second time lens of focal GDD $D_\mathrm{f}$ placed after the second dispersive medium, which creates a time telescope \cite{Srivastava23b}. 

Additional phase shift is not necessary in this interferometer and we set $\theta_1=0$. The even modes $\Psi_0$ and $\Psi_2$ are not affected by the action of the image inverter, $\mathcal{Z}_2A(t)=A(t)$, the transformation matrix in Eq. (\ref{Aout}) is a unit matrix for these modes, in other words, they experience a constructive interference with the field passing through the arm B of the interferometer and go ``right'' after the second beam splitter of the fist interferometer (Fig.~\ref{fig:interf}). The odd modes $\Psi_1$ and $\Psi_3$ acquire a $\pi$ phase shift in the image inverter, $\mathcal{Z}_2A(t)=-A(t)$, and, as a consequence, the transformation matrix in Eq. (\ref{Aout}) is antidiagonal for these modes, which means that they experience a destructive interference at the second beam splitter, and go ``down'' after this beam splitter.

We note that the time inversion realized by this interferometer does not mean the full temporal inversion of the photon wavefunction (impossible in a unitary system), because the time inversion is applied to the envelope only and not to the carrier wave.

\subsection{$Z_4$ interferometer}
The second interferometer separates the zeroth and second order modes and later the first and third order modes. This interferometer is equipped with a $Z_4$ gate in its arm A. For this device, $\gamma=-\pi/2$ and from Eqs.~(\ref{Dcondition}) and (\ref{DFcondition}) we have $D = D_\mathrm{f}=\tau^2$, which corresponds to a temporal Fourier processor. The field transformation in the temporal Fourier processor is given by Eq.~(\ref{transform}) with $a=d=0$ and $b=-1/c=-\tau^2$, supplemented by a phase shift by $\gamma/2=-\pi/4$:
\begin{equation}\label{transformFour}
   A(z,t) = \mathcal{Z}_4A(z',t) = \frac1{\sqrt{2\pi\tau^2}}\int e^{itt'/\tau^2}A(z',t')dt'. 
\end{equation}
This $Z_4$ gate, according to Eq.~(\ref{HGabcdFrac}), realizes a transformation $\Psi_n(t)\to e^{i\pi n/2}\Psi_n(t)$. Thus, the mode $\Psi_0$ is unaffected by this gate and goes ``right'' after the output beam splitter of the second interferometer (Fig.~\ref{fig:interf}). The mode $\Psi_2$ acquires a $\pi$ phase shift and goes ``down''. For these two modes we set $\theta_2=0$.

The modes $\Psi_1$ and $\Psi_3$ are delayed by $\Delta t\gg\tau$ and enter the second interferometer through the other port of its input beam splitter. During their passage an additional phase shift $\theta_2=\pi/2$ is added to the $Z_4$ gate, the overall phase shift being $\gamma/2+\theta_2=+\pi/4$. For this reason, the phase shifter in the second interferometer is denoted as ``variable'' in Fig.~\ref{fig:interf}. As a result, the mode $\Psi_1$ acquires a $\pi/2+\theta_2=\pi$ phase shift, experiences a destructive interference at the output beam splitter of the second interferometer and goes ``right'' after it. In contrast, the mode $\Psi_3$ acquires a $3\pi/2+\theta_2=2\pi$ phase shift, i.e. is unaffected, experiences a constructive interference at the output beam splitter of the second interferometer and goes ``down'' after it.

\subsection{Extension to higher dimensions} 
The cascade of two interferometers shown in Fig.~\ref{fig:interf} acts identically on modes of orders $n$ and $n+4$. The third interferometer may be added to separate the zeroth and fourth order modes and later other pairs of modes of orders $n$ and $n+4$. This interferometer should be equipped with a $Z_8$ gate in its arm A. For this device we have the fractional angle $\gamma=-\pi/4$ and, consequently, from Eqs.~(\ref{Dcondition}) and (\ref{DFcondition}), $D_\mathrm{f}=\sqrt{2}\tau^2$ and $D=D_\mathrm{f}(\sqrt{2}-1)/\sqrt{2}$, which corresponds to a temporal fractional Fourier processor of order $-\frac12$. The field transformation in this processor is given by Eq.~(\ref{transform}) with $a=d=\sqrt{2}/2$ and $b=-\sqrt{2}\pi\tau^2$, supplemented by a phase shift by $-\pi/8$:
\begin{equation}\label{transformFrac}
   A(z,t) = \frac{e^{i\pi/8}}{2^{1/4}\sqrt{\pi\tau^2}}\int e^{i\left(2^{-1/2}tt'-t^2-{t'}^2\right)/\tau^2}A(z',t')dt'. 
\end{equation}
This $Z_8$ gate, according to Eq.~(\ref{HGabcdFrac}), realizes a transformation $\Psi_n(t)\to e^{i\pi n/4}\Psi_n(t)$. The mode $\Psi_0$ passes it unaffected and exits ``right'' the third interferometer. The mode $\Psi_4$ propagating together with $\Psi_0$ acquires a phase $\pi$ and exits it ``down.'' After the time interval $\Delta t$ the modes $\Psi_1$ and $\Psi_5$ enter the third interferometer. During their passage, an additional phase shift $\theta_3=-\pi/4$ should be applied to the variable phase shifter in this interferometer, so that they acquire phase shifts $0$ and $\pi$ respectively and go in different directions after the output beam splitter of this interferometer.

The modes $\Psi_2$ and $\Psi_3$ should enter the third interferometer upon passing a delay line introducing the delay $2\Delta t$. They are separated from the accompanying modes $\Psi_6$ and $\Psi_7$ in a similar way.

\section{Sorting the temporal modes of SPDC} 
\begin{figure*}[!ht]
\centering
\includegraphics[width=\linewidth]{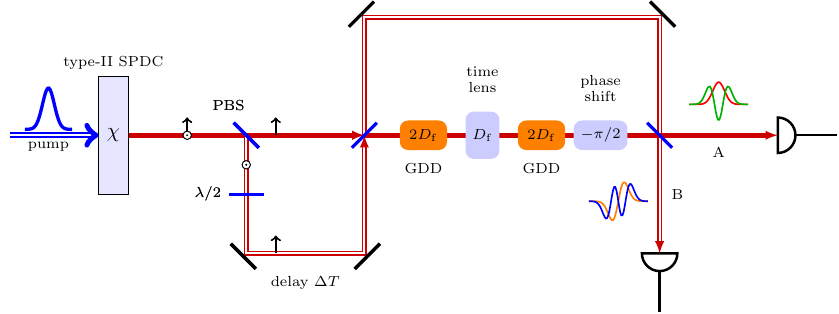}
\caption{Schematic of proposed experiment, where photon pairs are generated in frequency-degenerate collinear type-II SPDC with symmetric group velocity matching. The ordinary and extraordinary photons are separated on a polarizing beam splitter (PBS). The ordinary photon enters the sorter in beam A (single line). The extraordinary photon has its polarization rotated by a $\lambda/2$ plate and is delayed by $\Delta T$, then enters the sorter in beam B (double line). At the interferometer output, the even modes of the ordinary photon exit in beam A, while its odd modes exit in beam B. For the extraordinary photon, the exits are switched, the even modes exit in beam B, while the odd ones in beam A. \label{fig:SPDC}}
\end{figure*}

As an application of the proposed temporal mode sorter, we consider sorting of temporal modes of two entangled photons (a biphoton) generated in SPDC. Entanglement of photons means that a set of discrete spatial, temporal and polarization modes (Schmidt modes) can be defined for each photon, which are perfectly correlated: If one photon is found in a given Schmidt mode, the other photon is found in the corresponding Schmidt mode \cite{Law00}. The total number of entangled modes is characterized by the Schmidt number, which can be very high for the general case of spatio-temporal entanglement \cite{Gatti12,Horoshko12}. Here, we consider a regime of type-II collinear and frequency-degenerate SPDC where the generated photons have orthogonal polarizations, occupy the same spatial mode and are entangled in their temporal degree of freedom only. In addition, the condition of symmetric group-velocity matching can be achieved by means of quasi-phase matching at a given pump wavelength \cite{Ansari18} providing two benefits: The generated photons have identical temporal Schmidt modes and the number of these modes can be varied by tuning the pump bandwidth. 
The Schmidt modes of one photon can be sorted, and subsequently, those of the other. Since the central frequencies and modal shapes coincide for both photons, the same mode sorter can be applied to both of them. The scheme of the proposed experiment is shown in Fig. \ref{fig:SPDC}.

\subsection{Parametric downconversion \label{sec:PDC}}

We consider a nonlinear crystal with type-II frequency-degenerate phase matching. The plane-wave pump with the central frequency $\omega_p$ propagates along the $z$ axis and is polarized as either ordinary or extraordinary wave. As a result of the nonlinear transformation of the pump field in the crystal, two subharmonic waves with the central frequency  $\omega_0=\omega_p/2$ emerge, polarized as ordinary and extraordinary waves. The interaction of these waves is most easily described in the Heisenberg picture in the spectral domain. The positive-frequency part of the field of each wave (in photon-flux units) is
\begin{equation}\label{FourierSignal}
E_\mu^{(+)}(z,t) = \int  \epsilon_\mu(z,\Omega)e^{ik_\mu(\Omega)z-i(\omega_0+\Omega)t} \frac{d\Omega}{2\pi},		
\end{equation}
where $\mu$ takes values $\{p,o,e\}$ for the pump, ordinary and extraordinary subharmonic waves respectively, $t$ is time, $\Omega$ denotes the frequency detuning from the carrier frequency, and $k_\mu(\Omega)$ is the wave vector of the corresponding wave at frequency $\omega_\mu+\Omega$. The spectral amplitude of the pump $\epsilon_p(\Omega,z)$ is independent of $z$ since the pump is assumed to be undepleted. We assume also that it is a Gaussian transform-limited pulse with full width at half maximum (FWHM) duration $\tau_p$, whose peak passes through the position $z=0$ (crystal input face) at a time $t=0$, and write $\epsilon_p(z,\Omega) = \alpha(\Omega)=\alpha_0\exp(-\Omega^2/4\Omega_p^2)$, where $\Omega_p=\sqrt{2\ln 2}/\tau_p$.

The spectral amplitude of the ordinary or extraordinary subharmonic wave, $\epsilon_\mu(z,\Omega)$, is the annihilation operator of a photon at position $z$ with the frequency $\omega_\mu+\Omega$ and the corresponding polarization, satisfying the canonical equal-space commutation relations \cite{Huttner90,Kolobov99,Horoshko22} $\left[\epsilon_\mu(z,\Omega),\epsilon_\nu^\dagger(z,\Omega')\right]
= 2\pi\delta_{\mu\nu}\delta(\Omega-\Omega')$. The evolution of this operator along the crystal is described by the spatial Heisenberg equation \cite{Shen67}
\begin{equation}\label{evolution}
    \frac{d}{dz}\epsilon_\mu(z,\Omega) = \frac{i}\hbar\left[\epsilon_\mu(z,\Omega),G(z)\right],
\end{equation}
where the spatial Hamiltonian $G(z)$ is given by the momentum transferred through the plane $z$ \cite{Horoshko22} and equals
\begin{equation}\label{G}
    G(z) = \chi(z)\int\limits_{-\infty}^{+\infty} E^{(-)}_p(z,t)\hat E^{(+)}_o(z,t)\hat E^{(+)}_e(z,t)dt + \mathrm{H.c.},
\end{equation}
where $\chi(z)$ is the nonlinear coupling coefficient and $E^{(-)}_p(z,t)=E^{(+)*}_p(z,t)$ is the negative-frequency part of the field. In a periodically poled crystal, $\chi(z)$ changes its sign every distance of $\Lambda/2$, where $\Lambda$ is the poling period. i.e., represents a meander function. This function can be decomposed into Fourier series, where only the term of the order $-1$ affects the phase matching \cite{Horoshko17}. Thus, we write $\chi(z)\approx\chi_0\exp(-2\pi iz/\Lambda)$. Substituting Eqs.~(\ref{FourierSignal}) and (\ref{G}) into Eq.~(\ref{evolution}), performing the integration, and applying the commutation relations, we obtain the spatial evolution equations
\begin{equation}\label{evolution2}
\begin{split}
    \frac{d\epsilon_o(z,\Omega)}{dz} &=\kappa\int \alpha(\Omega+\Omega')\epsilon^{\dagger}_e(z,\Omega')e^{i\Delta(\Omega,\Omega')z}d\Omega',\\
    \frac{d\epsilon_e(z,\Omega)}{dz} &=\kappa\int \alpha(\Omega+\Omega')\epsilon^{\dagger}_o(z,\Omega')e^{i\Delta(\Omega',\Omega)z}d\Omega',
\end{split}
\end{equation}
where $\kappa=i\chi_0/2\pi\hbar$ is the new coupling constant and $\Delta(\Omega,\Omega')=k_p(\Omega+\Omega')-k_o(\Omega)-k_e(\Omega')+2\pi/\Lambda$ is the phase mismatch for the three interacting waves. 

In the low-gain regime, where the pump amplitude is sufficiently small, Eq.~(\ref{evolution2}) can be solved perturbatively, by substituting $\epsilon^{\dagger}_\mu(z,\Omega')\to\epsilon^{\dagger}_\mu(0,\Omega')$ under the integral. In this way, we obtain the fields at the output of a crystal of length $L$
\begin{equation}\label{CrystalSolution}
\begin{split}
    \epsilon_o(L,\Omega)&=\epsilon_o(0,\Omega) +\int    J(\Omega,\Omega')\epsilon^{\dagger}_e(0,\Omega') \frac{d\Omega'}{2\pi},\\  \epsilon_e(L,\Omega)&=\epsilon_e(0,\Omega) +\int   J(\Omega',\Omega)\epsilon^{\dagger}_o(0,\Omega') \frac{d\Omega'}{2\pi},
\end{split}
\end{equation}
where $J(\Omega,\Omega')=2\pi\kappa L\alpha(\Omega+\Omega')\Phi(\Omega,\Omega')$ is the joint spectral amplitude (JSA) of two generated photons and $\Phi(\Omega,\Omega')=e^{i\Delta(\Omega,\Omega')L/2}\sinc[\Delta(\Omega,\Omega')L/2]$ is the phase-matching function. 

A typical approximation in type-II SPDC consists in linearizing the dispesion law: $k_\mu(\Omega)\approx k_\mu^0+k_\mu'\Omega$, where $k_\mu^0=k_\mu(0)$ and $k_\mu'$ is the derivative of $k_\mu$ with respect to $\Omega$ at $\Omega=0$, having the meaning of inverse group velocity. In this approximation, the phase mismatch angle takes the form $\Delta(\Omega,\Omega')L/2\approx \tau_o\Omega+\tau_e\Omega'$, where $\tau_\mu=(k_p'-k_\mu')L/2$ and we assume that the poling period  $\Lambda$ is chosen so that the condition of phase matching at degeneracy $k_p^0-k_o^0-k_e^0+2\pi/\Lambda=0$ is satisfied. The time $\tau_\mu$ has the meaning of the interval by which the peak of the corresponding subharmonic pulse advances the peak of the pump pulse at the crystal output: The peak of the subharmonic is generated when the pump pulse is in the center of the crystal, the travel time for the pump and subharmonic peaks through the second half of the crystal are $k_p'L/2$ and $k_\mu'L/2$ respectively, hence the meaning of $\tau_\mu$ as the difference of these times. The JSA in this approximation reads 
\begin{eqnarray}\label{JSA}
J(\Omega,\Omega')&=&2\pi\kappa L\alpha_0e^{-\frac{(\Omega+\Omega')^2}{4\Omega_p^2} +i(\tau_o\Omega+\tau_e\Omega')}\\\nonumber
&&\times\sinc(\tau_o\Omega+\tau_e\Omega').
\end{eqnarray}

Substituting the solutions, Eq. (\ref{CrystalSolution}), into Eq. (\ref{FourierSignal}), we obtain the field transformation from the crystal input face to its output one. To write this transformation in a compact form, we define the group-delayed envelopes in the spirit of Eq. (\ref{Efield}):  $E_\mu^{(+)}(0,t)=A_{\mu,\text{vac}}(t) e^{-i\omega_0t}$ and $E_\mu^{(+)}(L,t)=A_\mu(t-k_\mu'L) e^{ik_\mu^0L-i\omega_0t}$. In this notation, the field transformation in the crystal has the form 
\begin{eqnarray}\label{BogoliubovA}
A_o(t) &=& A_{o,\text{vac}}(t)  + \int \tilde J(t,t') A_{e,\text{vac}}^\dagger(t')dt',\\\label{BogoliubovB}
A_e(t) &=& A_{e,\text{vac}}(t) + \int \tilde J(t',t) A_{o,\text{vac}}^\dagger(t')dt',
\end{eqnarray}
where the joint temporal amplitude (JTA) of two generated photons is the double Fourier transform of their JSA:
\begin{equation}\label{JtildeDef}
\tilde J(t,t') = \int\int J(\Omega,\Omega') e^{-i\Omega t-i\Omega't'}\frac{d\Omega d\Omega'}{(2\pi)^2}.
\end{equation}

Substituting Eq. (\ref{JSA}) into Eq. (\ref{JtildeDef}), we obtain
\begin{equation}\label{Jtilde}
\tilde J(t,t') 
= J_N \Pi\left(\frac{t-t'}{2\tau_-}-\frac12\right)e^{-\left(t\tau_e-t'\tau_o\right)^2\Omega_p^2/\tau_-^2},
\end{equation}
where $J_N=\sqrt{\pi}\kappa L\alpha_0\Omega_p/|\tau_-|$, $\Pi(x)$ is the rectangle function equal to 1 for $|x|<\frac12$ and 0 otherwise, and $\tau_-=\tau_o-\tau_e$. The same JTA was derived by considering the temporal evolution of the fields \cite{Keller97}, instead of the spatial evolution considered here.

\subsection{Gaussian modeling}
Both JSA and JTA can be significantly simplified by approximating them with double Gaussian functions. For this purpose, we replace the $\sinc(x)$ function in Eq. (\ref{JSA}) with the Gaussian function $e^{-x^2/2\sigma_s^2}$ having the same width at half maximum for $\sigma_s=1.61$ \cite{Grice01,Horoshko19}. Upon this replacement, JSA becomes a double Gaussian function. Substituting it into Eq. (\ref{JtildeDef}) and applying the multidimentional Gaussian integration technique \cite{Srivastava23,Srivastava23b}, we obtain JTA also in the form of a double Gaussian function
\begin{equation}\label{JtildeGauss}
\tilde J(t,t') 
= J_1 e^{-M_{11}(t-\tau_o)^2-M_{22}(t'-\tau_e)^2-2M_{12}(t-\tau_o)(t'-\tau_e)},
\end{equation}
where $J_1=2\kappa L\alpha_0\Omega_p^2/|T_o-T_e|$, $T_\mu=\sqrt{2}\Omega_p\tau_\mu/\sigma_s$, and $M_{ij}$ are elements of the matrix 
\begin{equation}\label{M}
    \mathbf{M} = \frac{\Omega_p^2}{(T_o-T_e)^2}\left(
    \begin{array}{cc}
        1+T_e^2 & -1-T_oT_e \\
        -1-T_oT_e & 1+T_o^2
    \end{array}
    \right).
\end{equation}

A double-Gaussian JTA can be represented in a form of Schmidt decomposition by means of the Mehler's formula for Hermite polynomials (see Appendix \ref{sec:appendix2}) 
\begin{equation}\label{Schmidt}
\tilde J(t,t') 
= \sqrt{P_b}\sum\limits_{n=0}^{\infty}\sqrt{\lambda_n}\psi_n(t)\varphi_n(t'),
\end{equation}
where $P_b=2\pi(\kappa L\alpha_0\Omega_p)^2/|T_o-T_e|$ is the probability of biphoton generation, $\psi_n(t)$ and $\varphi_n(t')$ are the Schmidt modal functions of the ordinary and extraordinary photons respectively, defined as
\begin{eqnarray}\label{psin}
\psi_n(t) &=& \frac1{\sqrt{\tau_1}}h_n\left(\frac{t-\tau_o}{\tau_1}\right),\\\label{phin}
\varphi_n(t) &=& \frac1{\sqrt{\tau_2}}h_n\left(\frac{t-\tau_e}{\tau_2}\right),
\end{eqnarray}
$\lambda_n$ are the Schmidt coefficients
\begin{equation}\label{lambda}
\lambda_n=\frac2{K+1}\left(\frac{K-1}{K+1}\right)^n  
\end{equation}
normalized to unity, $\sum_n\lambda_n=1$, and 
\begin{equation}\label{K}
K=\frac1{\sum_n\lambda_n^2} = \frac{\sqrt{(1+T_o^2)(1+T_e^2)}}{|T_o-T_e|}  
\end{equation}
is the Schmidt number. In the Schr\"odinger picture, the JTA gives the wave function of two photons, which are entangled when the JTA is not separable in its arguments. The Schmidt number $K$ shows the effective number of entangled modes for each photon and as such is a measure of the degree of entanglement \cite{Law04,Horoshko12,Gatti12}. The temporal widths of the Schmidt modes are 
\begin{equation}\label{tau12}
\tau_{1,2} = \frac{\sqrt{|T_o-T_e|}}{\sqrt{2}\Omega_p}\left(\frac{1+T_{o,e}^2}{1+T_{e,o}^2}\right)^\frac14.
\end{equation}

We see from Eqs. (\ref{psin}) and (\ref{phin}) that the modes with polarization $\mu=o,e$ are delayed by $\tau_\mu$. Indeed, in accord with the general formalism of Sec. II, the field envelopes at position $z=L$ are delayed by the group delay time $k_\mu'L$. However, the peak of the zeroth-order mode exits the crystal at time $t=k_\mu'L+\tau_\mu$ if the peak of the pump pulse entered it at time $t=0$: The peak of subharmonic is generated when the pump pulse is in the center of the crystal, which happens at time $t=k_p'L/2$, it passes the second half of the crystal during time $k_\mu'L/2$, and exits the crystal at time $t=k_p'L/2+k_\mu'L/2=k_\mu'L+\tau_\mu$. We see that the group delays for fields \emph{generated} in dispersive media are different from those for  fields \emph{passing} through them. 

Another observation from Eqs. (\ref{psin}) and (\ref{phin}) is that the temporal widths of the Schmidt modes are, in general, different for two photons. It means that the task of sorting these modes requires two different mode sorters, each tuned to a given temporal width $\tau_1$ or $\tau_2$. However, for applications like quantum cryptography or quantum computing with temporal modes \cite{Brecht15}, it is highly desirable to have one mode sorter applied subsequently to the first and then to the second photons of the pair. The next section shows how this mode symmetry can be reached.

\subsection{Symmetric group velocity matching \label{sec:symm}}
We see from Eq. (\ref{tau12}) that the practically interesting regime with $\tau_1=\tau_2$ can be reached when $\tau_o=\pm \tau_e$. In the case $\tau_o=\tau_e$, the kernels Eqs. (\ref{JSA}) and (\ref{Jtilde}) are not square-integrable and the linear approximation for the dispersion is invalid. The case $\tau_o=-\tau_e$ is more practical and is known as symmetric group velocity matching \cite{Keller97,Ansari18} or extended phase matching \cite{Giovannetti02}. It was first engineered in a crystal of periodically-poled potassium titanyl phosphate (ppKTP) \cite{Konig04}, which we consider here as an example. The pump propagates along the $X$ axis of this biaxial crystal and is polarized along the $Y$ axis, i.e. represents an ordinary wave. The two subharmonic waves are polarized along the $Y$ (ordinary wave) and $Z$ (extraordinary wave) axes of the crystal. The latter should not be confused with the $z$ axis of the reference frame, the direction of pump propagation.

The refractive indices for the $Y$ and $Z$ polarized waves are obtained from the Sellmeier equations for KTP \cite{Konig04} and, for a crystal of length $L=40$ mm pumped at wavelength $\lambda_p=2\pi c/\omega_p=791.5$ nm, we obtain $\tau_o=-\tau_e=2.95$ ps. The poling period required for reaching the phase matching at degeneracy is $\Lambda=47.6$ $\mu$m. Substituting $\tau_o=-\tau_e>0$ into Eq. (\ref{tau12}), we obtain
\begin{equation}\label{tau}
\tau_1 = \tau_2= \tau = \sqrt{\frac{\tau_o\tau_p}{\sigma_s\sqrt{\ln2}}} \approx 0.86\sqrt{\tau_o\tau_p}.
\end{equation}
The Schmidt number, Eq. (\ref{K}), takes the form 
\begin{equation}\label{Ksymm}
K= \frac12\left(T_o+\frac1{T_o}\right) = \frac12\left(\frac{\delta_s\tau_o}{\tau_p}+\frac{\tau_p}{\delta_s\tau_o}\right),
\end{equation}
where $\delta_s=2\sqrt{\ln2}/\sigma_s\approx1.03$. The minimal value of $K=1$ is reached at the pump pulse duration $\tau_p=\delta_s\tau_0$. In this case the photons are disentangled, each occupying just one Gaussian temporal mode, as required for some applications. We, however, are interested in a longer pump pulse, resulting in a moderate number of entangled modes. 

When choosing the pump pulse duration $\tau_p$, we need to remember that Eq. (\ref{TimeLens}) is valid only under the condition that the temporal aperture of a time lens surpasses the duration of the stretched pulse passing through it \cite{Kolner94}, otherwise including integrals over temporal point-spread functions \cite{Patera18}. 
Since we need to have the same carrier frequency before and after the lens, a four-wave mixing or electro-optical time lenses can be used. The temporal aperture of a parametric time lens based on four-wave mixing can be written as $T_A=D_\mathrm{f}\Omega_m$, where $\Omega_m=\sqrt{2}\pi \Delta f_\text{pump}$ and $\Delta f_\text{pump}$ is the FWHM of the intensity spectrum of the pump pulse \cite{Salem13}. This temporal aperture may be made sufficiently long for processing picosecond-long pulses generated in SPDC \cite{Joshi22}. However, a parametric time lens has typically a non-unit quantum conversion efficiency, which means a probabilistic action for single photons. A great advantage of a time lens based on electro-optical phase modulation consists in the deterministic nature of the electro-optical effect, providing an almost lossless operation \cite{Karpinski17,Sosnicki23}. Its main shortcoming is a rather short temporal aperture $T_A=\sqrt{\Phi D_\mathrm{f}}$ for the given focal GDD $D_\mathrm{f}$ and phase modulation amplitude $\Phi$ \cite{Kolner94,Salem13}. This shortcoming is removed in the recently proposed Fresnel time lens \cite{Sosnicki18,Sosnicki20}, successfully applied to bandwidth compression of single photons \cite{Sosnicki23}. The driving voltage in such a time lens is not sinusoidal, as in ordinary modulators \cite{Horoshko18}, but represents a wrapped parabola, created by an arbitrary waveform generator. To create a phase shift $\phi=t^2/2D_\mathrm{f}$ within the temporal aperture $T_A$, the modulator should realize the instantaneous circular frequency $\partial\phi/\partial t =t/D_\mathrm{f}$ at times $t=\pm T_A/2$, which is limited by the electronic bandwidth of the arbitrary waveform generator $f_\mathrm{RF}$, i.e. $T_A/2D_\mathrm{f}=2\pi f_\mathrm{RF}$. Thus, for a Fresnel time lens, the temporal aperture scales linearly with the focal GDD: $T_A=D_\mathrm{f}\Omega_m$, where $\Omega_m=4\pi f_\mathrm{RF}$. For a parametric or electro-optical Fresnel time lens, we write the condition on the time aperture as $T_1^F<D_\mathrm{f}\Omega_m$, where $T_1^F$ is the FWHM duration of the stretched pulse entering the time lens. Considering the zeroth-order Hermite-Gauss mode (a Gaussian pulse) of temporal scale $\tau$ undergoing a fractional Fourier transform, we arrive at the condition (Appendix \ref{sec:appendix1}) $\tau>\tau_G$, where $\tau_G = 4\sqrt{\ln2}/\Omega_m$. The spectral width of Hermite-Gauss modes scales approximately as $\sqrt{n}$ with the mode order $n$ \cite{LaVolpe21,Patera23} and we can estimate the duration of the stretched pulse by the same scaling factor. In order that $K$ modes fit into the temporal aperture, we need to satisfy the condition $\tau>\sqrt{K}\tau_G$. Substituting Eqs. (\ref{tau}) and (\ref{Ksymm}) (in the limit $\tau_p\gg\tau_o$) into this inequality, we obtain a limitation on the time lens bandwidth
\begin{equation}
    \Omega_m>\frac{c_m}{\tau_o},
\end{equation}
where $c_m=2\sqrt{\ln2}\sigma_s\approx2.68$. For an electrooptical Fresnel time lens, we obtain $f_\mathrm{RF}>72$ GHz, which is within the reach of modern technology, especially for integrated solutions \cite{Wang18}. Choosing $f_\mathrm{RF}=75$ GHz and $K=4$, we obtain $\tau_G=3.5$ ps, $\tau_p=24$ ps, and $\tau=7.3$ ps.

\subsection{Modulo $2$ (parity) sorting}

Now, we consider the field transformation in the temporal-image-inverting interferometer shown in Fig. \ref{fig:SPDC}. In the group-delayed reference frame of the ordinary wave, the field at the interferometer entrance is non-vacuum around time $\tau_o$, when the ordinary wave arrives, and around time $\tau_e+\Delta T$, when the extraordinary wave does. In the reference frame of the extraordinary wave, the group delay $\Delta T$ is included into the definition of the field envelope. We have thus two temporal windows of width $\Delta T$ each, the first centered at the peak of zeroth-order mode of the ordinary wave, the second at the peak of zeroth-order mode of the extraordinary one. Below, the window number will be indicated in square brackets in the upper index. We recall that the lower index is the interferometer number.

In the first window, we shift the time axis to the peak of the zeroth-order mode and define $A_1^{[1]\text{in}}(t)=A_o(t+\tau_o)$. The field at the other input is just vacuum $B_1^{[1]\text{in}}(t)=B_{1,\text{vac}}^{[1]}(t)$. The fields at the interferometer output are given by Eq. (\ref{Aout}), where $\ell=1$, $\theta_1=0$, and the action of operator $\mathcal{Z}_2$ is described by Eq. (\ref{transformInv}). The time lens in the interferometer is synchronized with the time axis, i.e. the vertex of the parabolic phase coincides with the passage of the peak of the zeroth-order mode; the effects of desynchronization can be found in Ref. \cite{Srivastava23}. At the interferometer output, we identify the detector of beam A as measuring parity $j=+1$ and define the field on its surface as $F_{+1}^{[1]}(t)=A_1^{[1]\text{out}}(t)$ and the detector of beam B as measuring parity $j=-1$ and define $F_{-1}^{[1]}(t)=B_1^{[1]\text{out}}(t)$. Applying the field transformation in the interferometer, Eq. (\ref{Aout}), we obtain (for $j=\pm1$)
\begin{equation}\label{Fj}
F_j^{[1]}(t) = \frac{\mathcal{Z}_2+j}2 A_1^{[1]\text{in}}(t) + \frac{\mathcal{Z}_2-j}2 B_{1,\text{vac}}^{[1]}(t).
\end{equation}

In the second window, we shift, in a similar way, the time axis to the peak of the zeroth-order mode and define $B_1^{[2]\text{in}}(t)=A_e(t+\tau_e)$. The field at the other input is just vacuum $A_1^{[2]\text{in}}(t)=A_{1,\text{vac}}^{[2]}(t)$. At the interferometer output, we identify the detector of beam B as measuring parity $k=+1$ and define $F_{+1}^{[2]}(t)=B_1^{[2]\text{out}}(t)$ the detector of beam A as measuring parity $k=-1$ and define  $F_{-1}^{[2]}(t)=A_1^{[1]\text{out}}(t)$. Applying the field transformation in the interferometer, as above, we obtain (for $k=\pm1$)
\begin{equation}\label{Fk}
F_k^{[2]}(t)=\frac{\mathcal{Z}_2+k}2 B_1^{[2]\text{in}}(t) + \frac{\mathcal{Z}_2-k}2 A_{1,\text{vac}}^{[2]}(t).
\end{equation}

\begin{figure}[!ht]
\centering
\includegraphics[width=\linewidth]{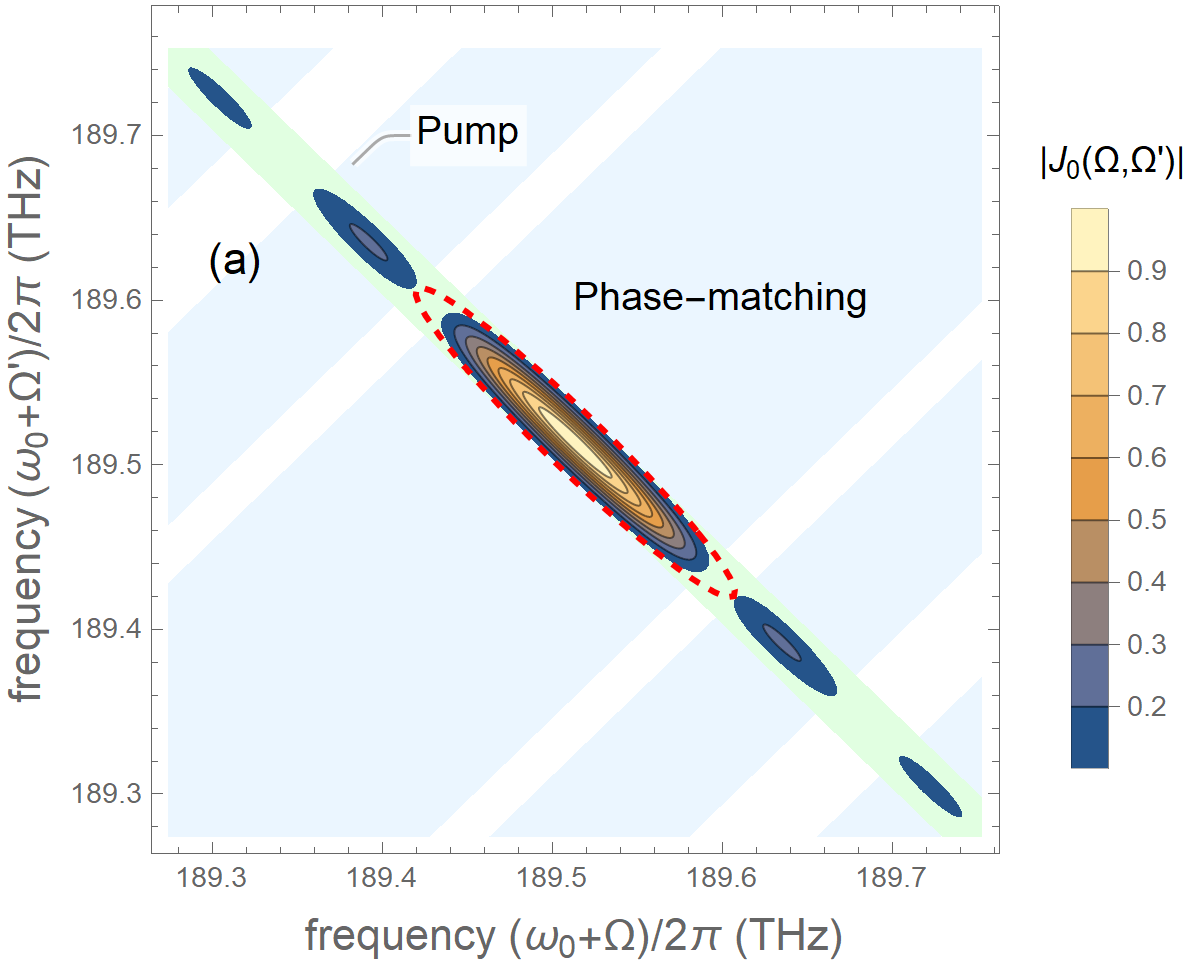}
\includegraphics[width=\linewidth]{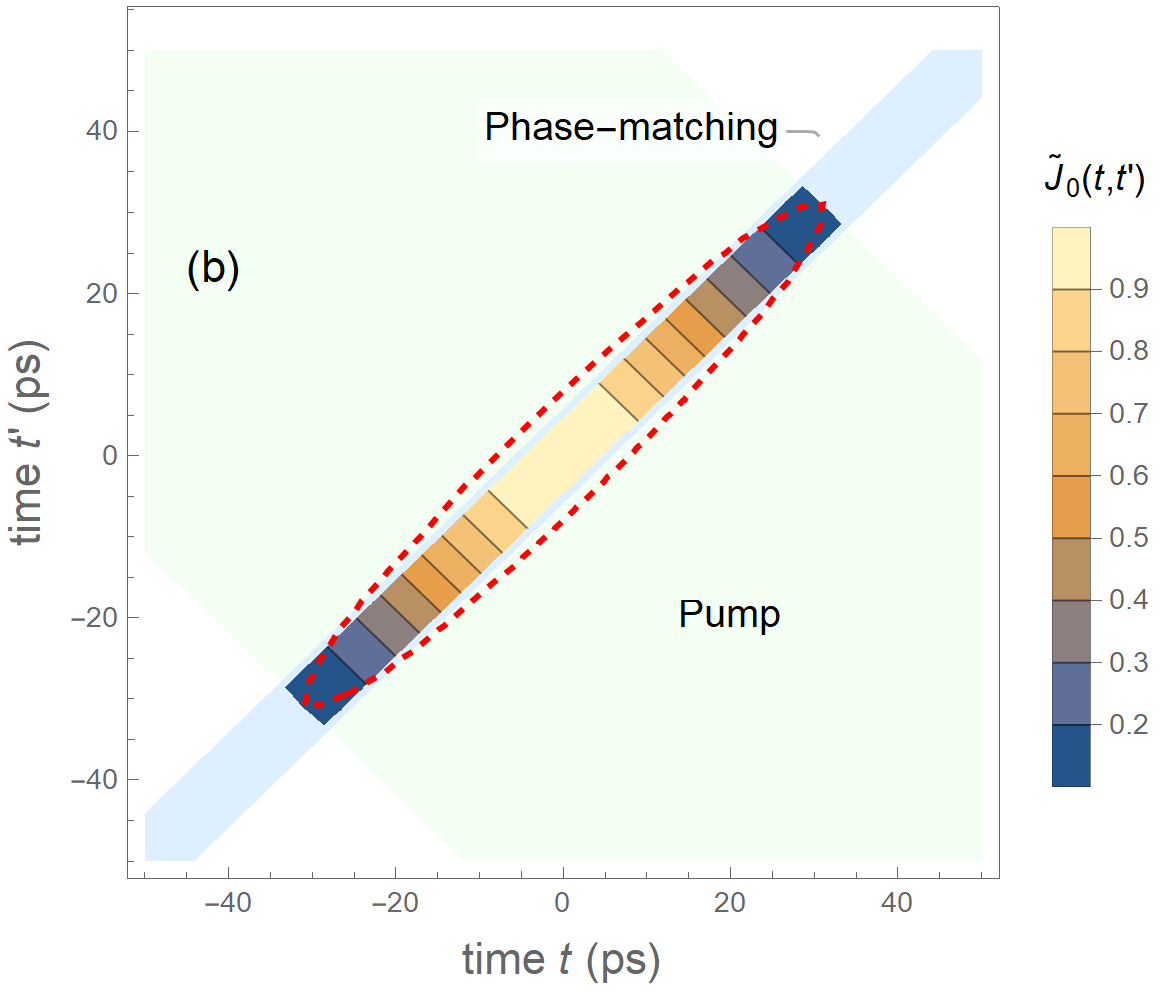}
\caption{(a) JSA of two photons generated in a 40-mm-long ppKTP crystal pumped at wavelength 791.5 nm by pulses of duration 24 ps. The JSA (color map) lies on the intersection of the pump-limited area and phase-matched area. The red dashed line shows the points where the Gaussian approximation is equal to 0.1. (b) JTA of the same photons lying on the intersection of the same two areas in the temporal domain. The red dashed line shows the points where the Gaussian approximation is equal to 0.1.\label{fig:JSA-JTA}}
\end{figure}

The probability to observe, in the same pumping cycle, one photon on detector $j=\pm1$ in the first window and another photon on detector $k=\pm1$ in the second window is 
\begin{equation}\label{Pjkdef}
P_{jk} = \int\limits_{-\Delta T/2}^{+\Delta T/2}  dt
\int\limits_{-\Delta T/2}^{+\Delta T/2} dt'
\left|C_{jk}(t,t') \right|^2,
\end{equation}
where $C_{jk}(t,t')=\langle F_j^{[1]}(t)F_k^{[2]}(t')\rangle$ is the anomalous correlator of the fields on the detectors. Here, we have used the Gaussian moment theorem \cite{Mandel-Wolf} and kept only the term of the lowest order in the smallness parameter of the low-gain regime of SPDC $\kappa L\alpha_0\Omega_p\ll1$ (see Appendix \ref{sec:appendix2}). Substituting Eqs. (\ref{BogoliubovA}), (\ref{BogoliubovB}), (\ref{Fj}), and (\ref{Fk}) into the anomalous correlator, applying commutation relations $[A(t),A^\dagger(t')]=\delta(t-t')$, and equating to zero all normally ordered correlators of vacuum fields, we obtain
\begin{equation}\label{Cjk}
C_{jk}(t,t') = \frac14\left(\mathcal{Z}_2+j\right)\left(\mathcal{Z}_2'+k\right)\tilde J_0(t,t'),
\end{equation}
where $\tilde J_0(t,t')=\tilde J(t+\tau_o,t'+\tau_e)$ is the zero-centered JTA of two generated photons and prime on the operator $\mathcal{Z}_2'$ means that it acts on a function of time $t'$. The zero-centered versions of the exact JTA, Eq. (\ref{Jtilde}) and its Gaussian approximation, Eq. (\ref{JtildeGauss}), for $\tau_o=-\tau_e>0$, are 
\begin{eqnarray}\label{J0E}
\tilde J_0^E(t,t') &=& \left(\frac\pi2\right)^\frac14\sqrt{\frac{P_b'\Omega_p}{4\tau_0}} \Pi\left(\frac{t-t'}{4\tau_o}\right)e^{-(t+t')^2\Omega_p^2/4},\\\label{J0G}
\tilde J_0^G(t,t') &=& \sqrt{\frac{P_b\Omega_p^2}{\pi T_0}} e^{-M_{11}(t^2+t'^2)-2M_{12}tt'},
\end{eqnarray}
where $P_b'=2(\pi/2)^{3/2}(\kappa L\alpha_0)^2\Omega_p/\tau_o$ is the biphoton generation probability calculated from the exact JTA, and the elements of the matrix $\mathbf{M}$ are $M_{11}=\Omega_p^2(T_o^2+1)/4T_o^2$ and $M_{12}=\Omega_p^2(T_o^2-1)/4T_o^2$. These functions are double Fourier transforms of the phase-shifted JSA, $J_0(\Omega,\Omega')=J(\Omega,\Omega')e^{-i(\tau_o\Omega+\tau_e\Omega')}$, and are shown in Fig. \ref{fig:JSA-JTA}.

To simplify the calculations, we consider any square-integrable complex function $g(t,t')$ as a vector in the space of square-integrable functions of $t$ and $t'$ and denote it by $|g)$. Similar to as it was done for ket and bra vectors by Dirac \cite{Dirac-Book}, we introduce the conjugated space of vectors $(f|$ with a scalar product  
\begin{equation}\label{fg}
(f|g) = \int\limits_{-\infty}^{+\infty}  dt
\int\limits_{-\infty}^{+\infty} dt' f^*(t,t')g(t,t').
\end{equation}
For any operator $\mathcal{Q}$, its conjugate $\mathcal{Q}^\dagger$ is defined by $(f|\mathcal{Q}^\dagger|g)=(g|\mathcal{Q}|f)^*$. In these notations and in the limit of sufficiently long $\Delta T$, we rewrite Eq. (\ref{Pjkdef})  as $P_{jk} = (\tilde J_0|\mathcal{O}_{jk}|\tilde J_0)$, where
\begin{equation}\label{Ojk}
\mathcal{O}_{jk} = \frac1{16}\left(\mathcal{Z}_2^\dagger+j\right)\left(\mathcal{Z}_2^{'\dagger}+k\right)
\left(\mathcal{Z}_2+j\right)\left(\mathcal{Z}_2'+k\right).
\end{equation}
Physically, this corresponds to calculation of outcome probabilities as averages of the elements of a positive operator-valued measure (POVM) $\mathcal{O}_{jk}$ in a specific interaction picture, where $\tilde J_0(t,t')$ is the temporal wave function of the biphoton. In this picture, the temporal dependence of the wave function includes the interaction of the waves in the source, but all interaction in the sorter is described by the evolution of the field operators. 

The time inversion operator $\mathcal{Z}_2$ is unitary, $\mathcal{Z}_2^\dagger\mathcal{Z}_2=1$, and Hermitian, $\mathcal{Z}_2^\dagger=\mathcal{Z}_2$, which results in all $\mathcal{O}_{jk}$ summing up to unity, as should be for a POVM. In addition, in the subspace of central-symmetric functions having the property $g(-t,-t')=g(t,t')$, to which $\tilde J_0(t,t')$ belongs, the following equalities hold: $\mathcal{Z}_2\mathcal{Z}_2'|g)=|g)$ and $\mathcal{Z}_2|g)=\mathcal{Z}_2'|g)$. From these properties and Eq. (\ref{Ojk}), we find $\mathcal{O}_{\pm1\mp1}=0$ and $\mathcal{O}_{\pm1\pm1}=(1\pm\mathcal{Z}_2)/2$. Finally, we obtain the probabilities of outcomes as
\begin{equation}\label{Pjk}
  P_{jk} = \frac{P_b}2\left(1+j\frac{(\tilde J_0|\mathcal{Z}_2|\tilde J_0)}{P_b}\right)\delta_{jk}.
\end{equation}
We see that the probability of detecting the photons with different parities is zero identically, even for the exact JTA. Thus, the theoretical cross-talk of the parity sorter is zero, which happens because the zero-centered JTA is central-symmetric on the $(t,t')$ plane. For the Gaussian model, we obtain from Eq. (\ref{J0G}) $(\tilde J_0^G|\mathcal{Z}_2|\tilde J_0^G)=P_b/K$. The probability of detecting an even mode, $(1+K^{-1})/2$ is higher than that of detecting an odd mode, $(1-K^{-1})/2$, however, with the growing number of modes, both probabilities approach $\frac12$. The fundamental cause of the advantage of even modes is the decreasing mode weight (Schmidt coefficient) with the mode number, determined by Eq. (\ref{lambda}). The weight of mode 1 is thus lower than that of mode 0. Similarly, the weight of mode 3 is lower than that of mode 2, et cetera. Consequently, the weight of all odd modes is lower than that of all even modes. These weights become asymptotically equal in the limit of infinite Schmidt number, where the weights of individual modes tend to a constant.

\subsection{Modulo $4$ sorting}
Now, we consider a sorter including two interferometers $\ell=1$ and $\ell=2$, as shown in Fig. \ref{fig:interf}.  The delay time between the interferometers is $\Delta t=\Delta T/2$, and therefore each time window of duration $\Delta T$ discussed in the preceding section is split into two sub-windows of durations $\Delta t$ so that the field is non-vacuum in the centers of these four sub-windows only. The sub-window number will be indicated in parenthesis in the upper index of the field envelope.

In the first sub-window of the first window, the input of the second interferometer is $A_2^{[1](1)\text{in}}(t)=A_1^{[1]\text{out}}(t)$ and $B_2^{[1](1)\text{in}}(t)=B_{2,\text{vac}}^{[1](1)}(t)$, where the latter field is vacuum. Its output fields are given, as usual, by Eq. (\ref{Aout}). The fields on the detectors are denoted by two sets of indices. Indices $u$ (for the first window) and $v$ (for the second window) run from 0 to 3 and denote the order of the mode arriving on this detector. Indices $j$, $k$, $l$, and $m$ take values $\pm1$ and denote the constructive (+1) or destructive (-1) interference in the first interferometer ($j$ for the first window, $k$ for the second one) and the second interferometer ($l$ for the first window, $m$ for the second one). The interference is constructive if the imaged photon remains in the same beam and destructive otherwise. Note, that $j$ and $k$ have the same meaning as in the preceding section. Thus, we write the field on the detector of beam A as $F_{0}^{[1]}(t)=F_{+1+1}^{[1]}(t)=A_2^{[1](1)\text{out}}(t)$ and that on the detector of beam B as $F_2^{[1]}(t)=F_{+1-1}^{[1]}(t)=B_2^{[1](1)\text{out}}(t)$. 

In the second sub-window of the first window, the input of the second interferometer is $A_2^{[1](2)\text{in}}(t)=A_{2,\text{vac}}^{[1](2)}(t)$ and $B_2^{[1](2)\text{in}}(t)=B_1^{[1]\text{out}}(t)$, where the former field is vacuum. In the output, we write the field on the detector of beam A as $F_1^{[1]}(t)=F_{-1-1}^{[1]}(t)=A_2^{[1](2)\text{out}}(t)$ and that on the detector of beam B as $F_3^{[1]}(t)=F_{-1+1}^{[1]}(t)=B_2^{[1](2)\text{out}}(t)$. 

In the first sub-window of the second window, the input of the second interferometer is $A_2^{[2](1)\text{in}}(t)=A_1^{[2]\text{out}}(t)$ and $B_2^{[2](1)\text{in}}(t)=B_{2,\text{vac}}^{[2](1)}(t)$, where the latter field is vacuum. In the output, we write the field on the detector of beam A as $F_3^{[2]}(t)=F_{-1+1}^{[2]}(t)=A_2^{[2](1)\text{out}}(t)$ and that on the detector of beam B as $F_1^{[2]}(t)=F_{-1-1}^{[2]}(t)=B_2^{[2](1)\text{out}}(t)$. 

Finally, in the second sub-window of the second window, the input of the second interferometer is $A_2^{[2](2)\text{in}}(t)=A_{2,\text{vac}}^{[2](2)}(t)$ and $B_2^{[2](2)\text{in}}(t)=B_1^{[2]\text{out}}(t)$, where the former field is vacuum. In the output, we write the field on the detector of beam A as $F_2^{[2]}(t)=F_{+1-1}^{[2]}(t)=A_2^{[2](2)\text{out}}(t)$ and that on the detector of beam B as $F_0^{[2]}(t)=F_{+1+1}^{[2]}(t)=B_2^{[2](2)\text{out}}(t)$. We see, that the values of $u=(0,1,2,3)$ correspond to values of $jl=($+1+1$,-1-1,+1-1,-1+1)$ respectively, with a similar relation between $v$ and $km$.

The probability of observing one photon in mode $u$ in the first window and another photon in mode $v$ in the second window is given by Eq. (\ref{Pjkdef}), where the indices should be changed as $jk\to uv=jl,km$. Substituting Eqs. (\ref{BogoliubovA}), (\ref{BogoliubovB}),  and (\ref{Aout}) into the anomalous correlator, applying the commutation relations, and equating to zero all normally ordered correlators of vacuum fields, as above, we write this probability as $P_{uv} = (\tilde J_0|\mathcal{O}_{uv}|\tilde J_0)$, where $\mathcal{O}_{uv}=\mathcal{Q}_{uv}^\dagger\mathcal{Q}_{uv}$ and
\begin{equation}\label{Quv}
\mathcal{Q}_{uv} = \frac1{16}\left(\mathcal{Z}_2+j\right)
\left(\mathcal{Z}_2'+k\right)
\left(\rho_j\mathcal{Z}_4+l\right)
\left(\rho_k\mathcal{Z}_4'+m\right),
\end{equation}
where the variable $\rho_j$ describes the additional phase shift $\theta_2=\pi/2$ applied to the odd modes and takes the values $\rho_{+1}=1$, $\rho_{-1}=i$. 

We see from the property $(\mathcal{Z}_2\pm1)(\mathcal{Z}_2'\mp1)|\tilde J_0)=0$ and Eq. (\ref{Quv}), that $P_{uv}=0$ when the parities of $u$ and $v$ differ, which is a consequence of errorless parity sorting in the first interferometer, as found in the preceding section. The only non-zero off-diagonal elements of the matrix $P_{uv}$ are $P_{02}=P_{20}$ and $P_{13}=P_{31}$, which describe the cross-talk between the corresponding modes of the same parity. To calculate these probabilities, we apply the properties of the operator $\mathcal{Z}_2$, established in the preceding section, and the following properties of the operator $\mathcal{Z}_4$: unitarity, $\mathcal{Z}_4^\dagger\mathcal{Z}_4=1$, commutativity with $\mathcal{Z}_2$, $[\mathcal{Z}_2,\mathcal{Z}_4]=[\mathcal{Z}_2^\dagger,\mathcal{Z}_4]=0$, and the property $\mathcal{Z}_4^\dagger=\mathcal{Z}_2\mathcal{Z}_4$. These properties are naturally obtained when $\mathcal{Z}_2$ and $\mathcal{Z}_4$ are represented as rotations by angles $\pi$ and $\pi/2$ respectively in the chronocyclic space of harmonic oscillator \cite{Karpinski21}. In addition, we employ the symmetricity of the zero-centered JTA with respect to the exchange of its arguments, $\tilde J_0(t',t)=\tilde J_0(t,t')$, which gives $(\tilde J_0|\mathcal{Z}_4|\tilde J_0)=(\tilde J_0|\mathcal{Z}_4'|\tilde J_0)$. From all these properties and Eq. (\ref{Quv}), we find $P_{02}=P_{20}=P_b'(p_1+p_2)$ and $P_{13}=P_{31}=P_b'(p_1-p_2)$, where
\begin{eqnarray}\label{p1def}
p_1 &=& \frac1{8P_b'}(\tilde J_0|1-\mathcal{Z}_2\mathcal{Z}_4\mathcal{Z}_4'|\tilde J_0),\\\label{p2def}
p_2 &=& \frac1{8P_b'}(\tilde J_0|\mathcal{Z}_2-\mathcal{Z}_4\mathcal{Z}_4'|\tilde J_0).
\end{eqnarray}

For the Gaussian JTA given by Eq. (\ref{J0G}), $p_1=p_2=0$ and the sorter is errorless. Physically, it means that a Fourier-transformed mode overlaps perfectly with itself at the output beam splitter of the second interferometer, since a Hermite-Gauss function is an eigenfunction of the Fourier transform. For the exact JTA given by Eq. (\ref{J0E}), however, these probabilities are non-zero, because the Schmidt modes are slightly different from Hermite-Gauss modes and are not eigenfunctions of the Fourier transform exactly. Substituting Eq. (\ref{J0E}) into Eqs. (\ref{p1def}) and (\ref{p2def}) and performing the integration, we find
\begin{eqnarray}\label{p1}
p_1 &=& \frac18\left[1-\sqrt{\pi\ln2}\frac{\tau^2}{\tau_o\tau_p} \erf^2\left(\frac{\tau_o\tau_p}{\sqrt{2\ln2}\tau^2}\right)\right],\\\label{p2}
p_2 &=& \frac1{8}\left[\sqrt{\frac\pi{4\ln2}}\frac{\tau_p}{\tau_o} \erf^2\left(\frac{\sqrt{2\ln2}\tau_o}{\tau_p}\right)\right.\\\nonumber
&&-\left.\sqrt{\frac{4\ln2}\pi}\frac{\tau^2}{\tau_o\tau_p\sqrt{1+(2\ln2)^2\tau_o^4/\tau_p^4}} \Si\left(\frac{2\tau_o^2}{\tau^2}\right)\right],
\end{eqnarray}
where $\erf(x)$ and $\Si(x)$ are the error function and the sine integral, respectively.

\begin{figure}[!ht]
\centering
\includegraphics[width=\linewidth]{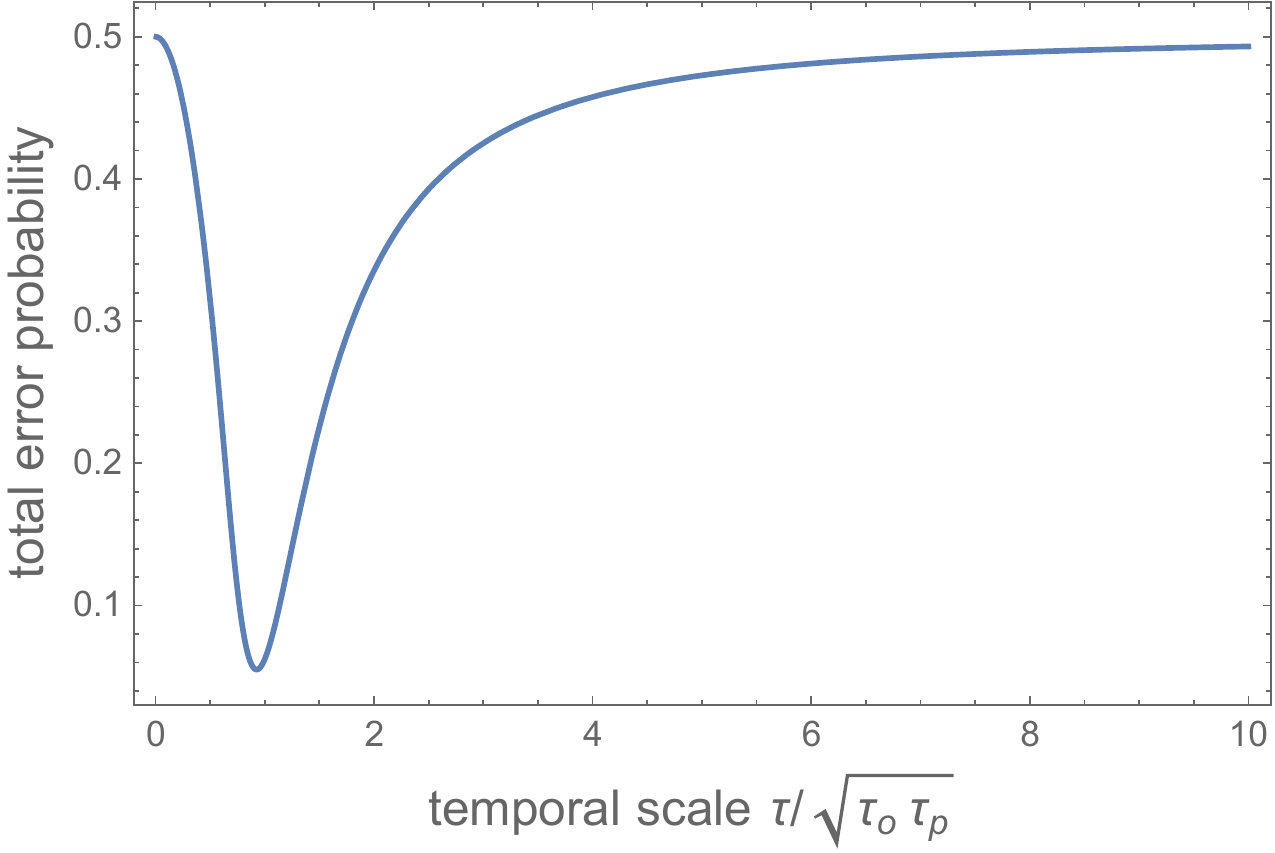}
\caption{Total probability of error $p_\text{tot}$ as a function of the modulo 4 sorter temporal scale $\tau$.\label{fig:ptot}}
\end{figure}

The total conditional probability of error under condition that a biphoton is generated is $p_\text{tot}=(P_{02}+P_{20}+P_{13}+P_{31})/P_b'=4p_1$. We see from Eq. (\ref{p1}) that this probability is a function of just one argument, $\tau/\sqrt{\tau_o\tau_p}$. This functional dependence is shown in Fig. \ref{fig:ptot}.

The minimal value of $p_\text{tot}=0.055$ is reached at $\tau=0.93\sqrt{\tau_o\tau_p}$, which is slightly different from the optimal value of $\tau$ given by Eq. (\ref{tau}) for the Gaussian model. Thus, the minimal total cross-talk error of a modulo 4 sorter is 5.5\%, which value is determined by the maximal overlap of a rectangle function with a Gaussian.

\begin{figure}[!ht]
\centering
\includegraphics[width=\linewidth]{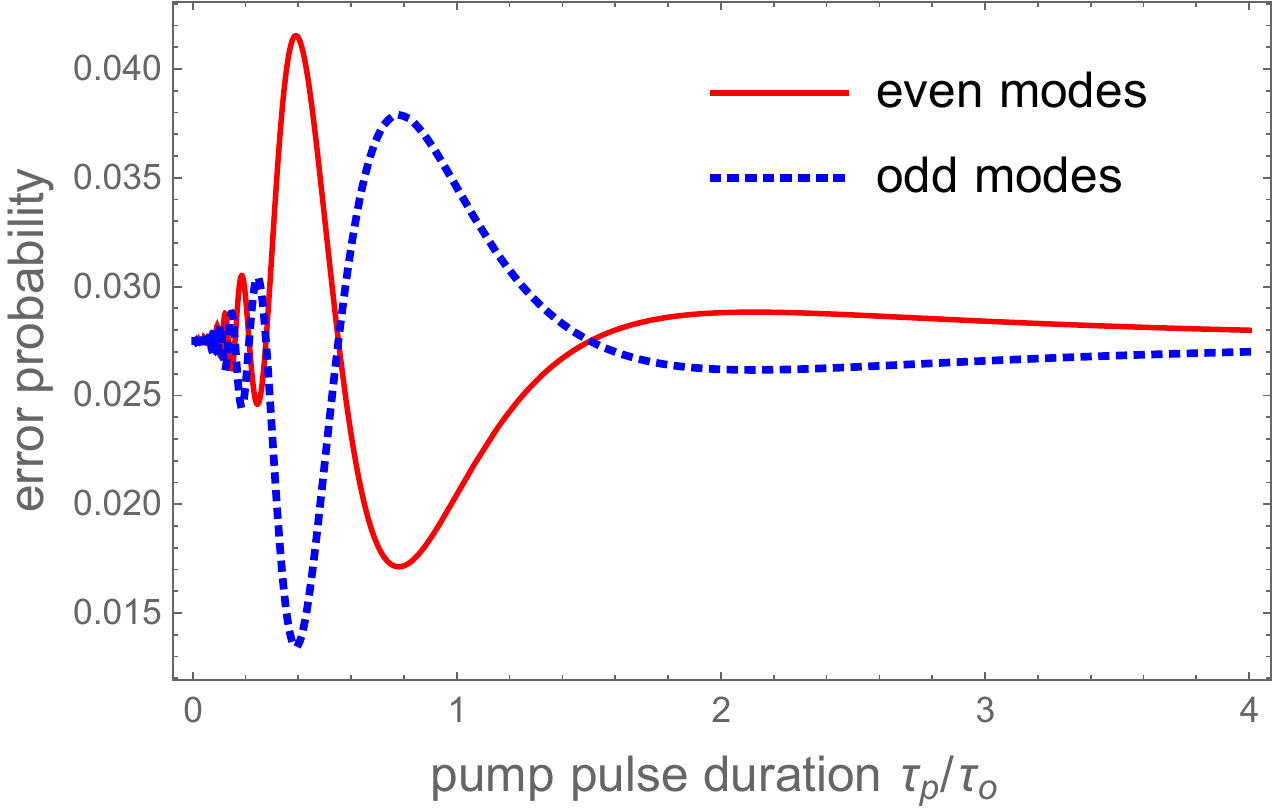}
\caption{Probabilities of error for even and odd modes as functions of the pump pulse duration $\tau_p$, when the optimal temporal scale $\tau$ is used in the modulo 4 sorter.\label{fig:peven}}
\end{figure}

Substituting the optimal value of the temporal scale $\tau$ into Eq. (\ref{p2}), we obtain $p_2$ as a function of the ratio $\tau_p/\tau_o$. This allows us to calculate conditional probabilities of errors in even modes $p_\text{even}= (P_{02}+P_{20})/P_b'=2(p_1+p_2)$ and in odd modes $p_\text{odd}= (P_{13}+P_{31})/P_b'=2(p_1-p_2)$, which sum up to the total error probability (Fig. \ref{fig:peven}). We see that for $\tau_p\approx8\tau_o$, as required for the choice $K=4$, $p_2$ tends to zero and the probability of error is almost equally distributed between the even and odd modes.

\section{Conclusion}
We have suggested a sorter of Hermite-Gauss temporal modes of light based on a cascade of Mach-Zehnder interferometers, each containing a time lens in one of its arms. We have shown how the temporal counterpart of the Gouy phase, which we call ``the temporal Gouy phase'', allows one to obtain constructive or desctructive interference of modes of different orders, and, as a consequence, direct them to different interferometer outputs. We have considered in detail the operation of a parity sorter and a modulo 4 sorter, applied to mode sorting of two entangled photons generated by SPDC in a nonlinear crystal with symmetric group velocity matching. For the parity sorter, we have shown that its theoretical cross-talk probability is zero, even when the eigenmodes of the source are not Hermite-Gauss modes exactly; it is sufficient that the JTA is central-symmetric. For the modulo 4 sorter, we have found the theoretical total probability of error on the level of 5.5\%. If this value of error is too high for a quantum network, e.g. for quantum error correction in linear quantum computation, then apodization should be applied to the SPDC crystal, which allows one to obtain a Gaussian phase-matching function \cite{Dixon13}. 

The proposed mode sorter can be applied to classical fields for manipulating the temporal modal structure of light pulses. It can also be applied to fragile quantum states of light, especially if realized on a single photonic chip with low loss \cite{Wang18}. Such a realization will open up bright perspectives for quantum information processing and quantum sensing with high-dimensional quantum encoding based on temporal modes of light \cite{Brecht15,Raymer20,Karpinski21}.

\section*{Acknowledgments} 
D. B. H. thanks Nicolas Treps for an insightful discussion. This work is supported by the network QuantERA of the European Union’s Horizon 2020 research and innovation programme under project “Quantum information and communication with high-dimensional encoding” (QuICHE) and funded by Agence Nationale de la Recherche, France, grant ANR-19-QUANT-0001.

\appendix
\section{Transformation of a Hermite-Gauss function under a linear canonical transform \label{sec:appendix0}}

The generating function for the Hermite polynomials reads
\begin{equation}\label{genfunc}
e^{2x\xi-\xi^2} = \sum\limits_{n=0}^\infty H_n(x)\frac{\xi^n}{n!},
\end{equation}
therefore, the Hermite-Gauss function can be written as 
\begin{equation}
\Psi_n(t) = \left(\tau 2^nn!\sqrt{\pi}\right)^{-\frac12}\left.\frac{\partial^n}{\partial\xi^n}\right|_{\xi=0} e^{2t\xi/\tau-\xi^2-t^2/2\tau^2}.
\end{equation}
After a linear canonical transform with the kernel defined by Eq. (\ref{KT}), this function transforms into
\begin{eqnarray}
\Psi_n^\text{out}(t) &=& \int K_\mathbf{T}(t,t')\Psi_n(t')dt' = \left(\tau 2^nn!\sqrt{\pi}\right)^{-\frac12}\\\nonumber
&\times& \left.\frac{\partial^n}{\partial\xi^n}\right|_{\xi=0} \int K_\mathbf{T}(t,t')e^{2t'\xi/\tau-\xi^2-t^{'2}/2\tau^2}dt'.
\end{eqnarray}

Taking the Gaussian integral over $t'$, we obtain
\begin{eqnarray}\nonumber
\Psi_n^\text{out}(t) &=& \left(\tau\beta 2^nn!\sqrt{\pi}\right)^{-\frac12}e^{-i\gamma/2-\alpha t^2/2\tau^2-t^2/2\tau^2\beta^2}\\
&\times& \left.\frac{\partial^n}{\partial\xi^n}\right|_{\xi=0} e^{2t\tilde\xi/\tau\beta-{\tilde\xi}^2},
\end{eqnarray}
where $\tilde\xi=\xi e^{-i\gamma}$. Substituting Eq. (\ref{genfunc}) with $\xi$ replaced by $\tilde\xi$ and taking the $n$th derivative over $\xi$, we obtain
\begin{eqnarray}\nonumber
\Psi_n^\text{out}(t) &=& \left(\tau\beta 2^nn!\sqrt{\pi}\right)^{-\frac12}e^{-i\gamma/2-\alpha t^2/2\tau^2-t^2/2\tau^2\beta^2}\\
&\times& H_n(t/\tau\beta) e^{-in\gamma},
\end{eqnarray}
which coincides with the right-hand side of Eq. (\ref{HGabcd}).

\section{Passage of a Gaussian pulse through a fractional Fourier processor \label{sec:appendix1}}

We consider a classical field with the envelope $Y_0(t)=E_0e^{-t^2/4\Delta t_0^2}$, where $E_0$ is the peak amplitude and $\Delta t_0$ is the intensity standard deviation.
In the frequency domain, we have
\begin{equation}
    \tilde Y_0(\Omega) = \int Y_0(t) e^{i\Omega t}dt = 2\sqrt{\pi}\Delta t_0E_0e^{-\Delta t_0^2\Omega^2},
\end{equation}
so that the standard deviation of the intensity spectrum $S_0(\Omega) = |\tilde Y_0(\Omega)|^2$ is $\Delta\Omega_0=1/2\Delta t_0$. For the FWHM temporal and spectral widths $T_0^\mathrm{F}=2\sqrt{2\ln2}\Delta t_0$ and $\Omega_0^\mathrm{F}=2\sqrt{2\ln2}\Delta\Omega_0$ respectively, we find the time-bandwidth product $T_0^\mathrm{F}\Omega_0^\mathrm{F} = 4\ln2\approx 2\pi\times0.44$, as expected for a Fourier-limited Gaussian pulse.

This pulse passes through a dispersive medium of GDD $D_\mathrm{in}$. In the group-delayed reference frame, defined by Eq. (\ref{Efield}), the field at the output is $\tilde Y_1(\Omega) = \tilde Y_0(\Omega)e^{iD_\mathrm{in}\Omega^2/2}$, which gives in the time domain
\begin{equation}
    Y_1(t) = \int \tilde Y_1(\Omega) e^{-i\Omega t}\frac{d\Omega}{2\pi}
    = E_1e^{-t^2/4\Delta t_1^2 - it^2/2C_1},
\end{equation}
where $E_1=E_0/\sqrt{1-iD_\mathrm{in}/2\Delta t_0^2}$ is the new peak amplitude, $C_1 = D_\mathrm{in}+4 \Delta t_0^4/D_\mathrm{in}$ is the chirp coefficient of the output pulse, and
\begin{equation}\label{Deltat1}
    \Delta t_1 = \sqrt{\Delta t_0^2+D_\mathrm{in}^2/4\Delta t_0^2}
\end{equation}
is the intensity standard deviation of the temporal distribution of the output pulse. The spectral distribution is unchanged $|\tilde Y_1(\Omega)|^2 = |\tilde Y_0(\Omega)|^2$ and the spectral standard deviation is $\Delta\Omega_1=\Delta\Omega_0$.

Then, the stretched pulse passes through a time lens and its envelope becomes $Y_2(t)=e^{it^2/2D_\mathrm{f}}Y_1(t)$ in accord with Eq. (\ref{TimeLens}). In the frequency domain, it gives
\begin{equation}
\tilde Y_2(\Omega) = \int Y_2(t) e^{i\Omega t}dt
= \tilde E_2e^{-\Omega^2/4\Delta\Omega_2^2 + iD_2\Omega^2/2},
\end{equation}
where $\tilde E_2=2\sqrt{\pi}\Delta t_1E_1[1+2i\Delta t_1^2/C_2]^{-1/2}$ is the spectral peak amplitude, $C_2^{-1}=C_1^{-1}-D_\mathrm{f}^{-1}$, $D_2=[C_2/4\Delta t_1^4+1/C_2]^{-1}$ and 
\begin{equation}\label{DeltaOmega2}
\Delta\Omega_2 = \Delta\Omega_0\sqrt{(1-R)^2+r^2 R^2}
\end{equation}
is the spectral standard deviation with $R=D_\mathrm{in}/D_\mathrm{f}$ and $r=2\Delta t_0^2/D_\mathrm{in}$. For a fractional Fourier transform, where conditions (\ref{Dcondition}) and (\ref{DFcondition}) are satisfied for $\tau=\sqrt{2}\Delta t_0$, we substitute $R=1-\cos\gamma$, $r=-\sin\gamma/(1-\cos\gamma)$ and obtain $\Delta\Omega_2 = \Delta\Omega_0$. In a similar manner, we obtain $D_2=-D_\text{in}$. This chirp is compensated in the second dispersive medium. Thus, the fractional Fourier transform does not change the bandwidth of a pulse of duration $\tau$ and does not chirp it.

The condition on the temporal aperture from Sec. \ref{sec:symm} is $2\sqrt{2\ln2}\Delta t_1<D_\mathrm{f}\Omega_m$, which can be rewritten as
\begin{equation}
    4\ln2\frac{R^2(1+r^2)}{\tau^2}<\Omega_m^2,
\end{equation}
or, upon substitutions for $R$ and $r$ as above,
\begin{equation}
    \tau>2\sqrt{2\ln2}\frac{\sqrt{1-\cos\gamma}}{\Omega_m}.
\end{equation}
The most stringent condition on the mode scale is imposed by the first gate with $\gamma=-\pi$.

\section{Temporal Schmidt decomposition for type-II biphoton \label{sec:appendix2}}

Probability density for detecting at the crystal output an ordinary photon at time $t$ and an extraordinary one at time $t'$ is $\langle E_e^{(-)}(L,t')E_o^{(+)}(L,t)E_o^{(+)}(L,t)E_e^{(+)}(L,t')\rangle\approx|\langle E_o^{(+)}(L,t)E_e^{(+)}(L,t')\rangle|^2 =|\langle A_o(t)A_e(t')\rangle|^2$, where we have used the Gaussian moment theorem \cite{Mandel-Wolf} and kept only the term of the lowest order in the smallness parameter of the low-gain regime of SPDC $\kappa L\alpha_0\Omega_p\ll1$. Substituting Eqs. (\ref{BogoliubovA}) and (\ref{BogoliubovB}) into this probability density and integrating it over both times, we arrive at the total probability of biphoton generation
\begin{equation}
    P_b=\int\int |\tilde J(t,t')|^2dtdt'=\pi J_1^2\frac{|T_o-T_e|}{\Omega_p^2}
\end{equation}
where we have used the JTA defined by Eq. (\ref{JtildeGauss}). 

The spectral decomposition of a real symmetric double-Gaussian kernel is given by multiplying both sides of the Mehler's formula for Hermite polynomials by $e^{-x^2/2-y^2/2}$ \cite{Grice01,Horoshko19}
\begin{equation}\label{Mehler}
\frac1{\sqrt{\pi}}e^{-\frac{1+q^2}{2(1-q^2)}\left(x^2+y^2\right)+\frac{2q}{1-q^2}xy}
= \sum_{n=0}^\infty pq^n h_n(x) h_n(y),
\end{equation}
where $-1<q<1$, $p=\sqrt{1-q^2}$, and $h_n(x)$ are the Hermite-Gauss functions introduced in Sec. \ref{sec:HGmodes}. Identifying $x=(t-\tau_o)/\tau_1$, $y=(t'-\tau_e)/\tau_2$, $q=\sqrt{(K-1)/(K+1)}$, and equalizing (up to a constant factor) the right-hand side of Eq. (\ref{JtildeGauss}) to the left-hand side of Eq. (\ref{Mehler}), we find the following system of equations:
\begin{eqnarray}
\frac{K}{\tau_1^2} &=& 2\Omega_p^2\frac{1+T_e^2}{(T_o-T_e)^2},\\
\frac{K}{\tau_2^2} &=& 2\Omega_p^2\frac{1+T_o^2}{(T_o-T_e)^2},\\
\frac{\sqrt{K^2-1}}{\tau_1\tau_2} &=& 2\Omega_p^2\frac{1+T_oT_e}{(T_o-T_e)^2}.
\end{eqnarray}
Solving this system for $K$, $\tau_1$ and $\tau_2$, we obtain Eqs. (\ref{K}) and (\ref{tau12}). We imply here that $1+T_oT_e>0$, which is the regime of our interest in Sec. IV. 

The singular value (Schmidt) decomposition of JTA can be easily obtained from its spectral decomposition, especially in the case of a positive $q$, considered here. Since all eigenvalues $pq^n$ are positive, they are also the singular values. The corresponding singular functions are given by the eigenfunctions, Eqs. (\ref{psin}) and (\ref{phin}).

\bibliography{Modes2023}

\end{document}